\documentclass[preprint,review,12pt]{elsarticle}
\usepackage{times,amsmath,amssymb,enumerate,bm,graphicx}
\usepackage{color}
\usepackage{amsthm}

\newtheorem{Theorem}{Theorem}

\newtheorem{lemma}{Lemma}
\newtheorem{example}{Example}

\journal{Signal Processing}
\begin{document}
\begin{frontmatter}

\title{The Geometry of Fusion Inspired Channel Design}

\author[Affil1]{Yuan Wang\corref{cor1}}
\ead{wangy@stat.colostate.edu}
\author[Affil1]{Haonan Wang}
\ead{wanghn@stat.colostate.edu}
\author[Affil1,Affil2]{Louis L. Scharf}
\ead{scharf@engr.colostate.edu}
\address[Affil1]{Department of Statistics, Colorado State University, Fort Collins, CO 80523, USA.}
\address[Affil2]{Department of Mathematics, Colorado State University, Fort Collins, CO 80523, USA. }
\cortext[cor1]{Corresponding author. Tel.: +1 970-491-3778.}

\begin{abstract}
This paper is  motivated by the problem of integrating multiple sources of measurements. We consider  two multiple-input-multiple-output (MIMO) channels, a primary channel and a secondary channel, with dependent input signals. The primary channel carries the signal of interest, and the secondary channel carries a signal that shares a joint distribution with the primary signal.  The problem of particular interest is designing the secondary channel matrix, when the primary channel matrix is fixed. We formulate the problem as an optimization problem, in which the optimal secondary channel matrix maximizes an information-based criterion. An analytical solution is provided in a special case. Two fast-to-compute algorithms, one extrinsic and the other intrinsic, are proposed to approximate the optimal solutions in general cases. In particular, the intrinsic algorithm exploits the geometry of the unit sphere, a manifold embedded in Euclidean space. The performances of the proposed algorithms are examined through a simulation study. A discussion of the choice of dimension for the secondary channel is given.
\end{abstract}


\end{frontmatter}

\section{Introduction}
\begin{figure}[htbp]
\centering
\includegraphics[totalheight=2.5in]{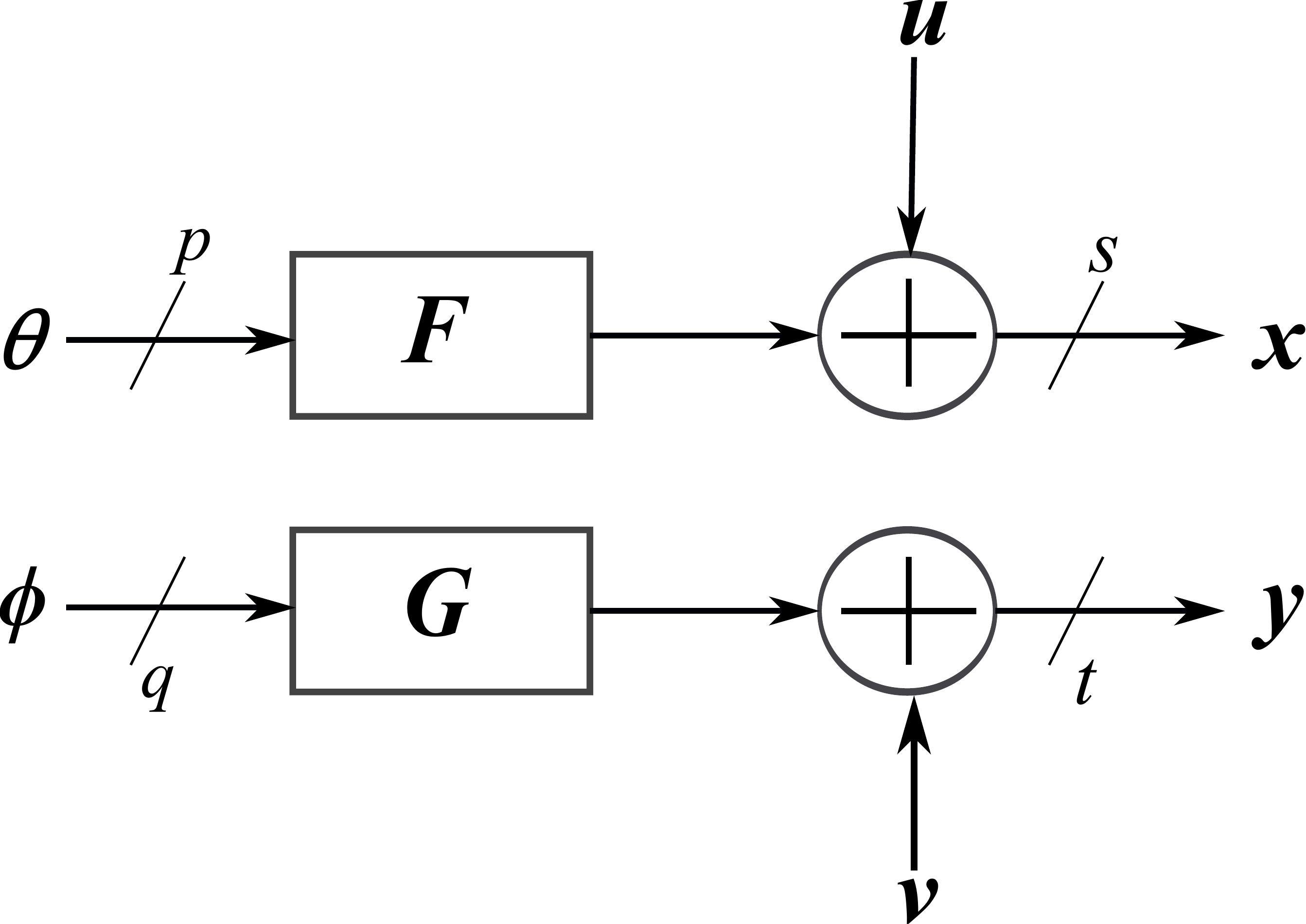}
\caption{A two-channel system with two linear channels.}
\label{fig:linsystem}
\end{figure}

Consider the following two-channel system, as illustrated in Fig.~\ref{fig:linsystem},
\begin{eqnarray}\label{eqn:linsys}
\begin{array}{l}
\bm x=\bm F\bm\theta+\bm{u} \\
\bm y=\bm G\bm\phi+\bm{v}.
\end{array}
\end{eqnarray}
The first channel is the primary channel that carries the signal of interest $\bm\theta$. The secondary channel carries a signal $\bm\phi$ that shares a joint distribution with $\bm\theta$.
The measurements $\bm x$ and $\bm y$ are linear transformations of the input signals with measurement noises  $\bm u$ and $\bm v$, respectively. For example, the elements of the primary signal $\bm\theta$ may be the complex scattering coefficients of several radar-scattering targets and the elements of the secondary signal $\bm \phi$  may be intensities in an optical map of these same optical-scattering targets. The measurement $\bm x$ is then a range-doppler map and the measurement $\bm y$ is an optical image.  We assume a known signal model, i.e., the joint distribution of $\bm\theta$ and $\bm\phi$. When the signals $\bm\theta$ and $\bm\phi$ are correlated, the measurements $\bm x$ and $\bm y$ both contain information about $\bm\theta$ and we can combine them to estimate $\bm\theta$. The fused estimate is expected to perform better than the estimate from a single source of measurements. In this paper, our objective is to design the channel matrix $\bm G$, with the primary channel fixed, such that the fused estimate achieves the best performance.

For a one-channel system $\bm x=\bm F\bm\theta+\bm u$, designing the channel matrix $\bm F$ exhibits parallels to the linear precoding design problem for multiple-input-multiple-output (MIMO) communication systems by considering $\bm F$ as the precoder into an identity channel matrix. The linear precoding design for MIMO channels has been studied in the literature \cite{Palomar:Cioffi:Lagunas:2003}-\cite{Palomar:Jiang:2006}. The optimal precoding is designed under various criteria, for example, signal-to-noise ratio (SNR) and signal-to-interference-noise ratio (SINR), \cite{Palomar:Cioffi:Lagunas:2003}-\cite{Scaglione:Stoica:Barbarossa:Giannakis:Sampath:2002}. Another criterion that has drawn more attention recently is the mutual information between input and output signals, \cite{Perezcruz:Rodrigues:Verdu:2010}-\cite{Carson:Calderbank}, \cite{Liu:Chong:Scharf:2012}-\cite{Lamarca:2009}. This information-based criterion is connected with estimation theory in a vector Gaussian channel with arbitrary input distribution by linking the mutual information with the minimum mean squared error (MMSE) \cite{Guo:Shamai:Verdu:2005}-\cite{Palormar:Verdu:2006}. In~\cite{Perezcruz:Rodrigues:Verdu:2010}, an optimal precoding matrix for the MIMO Gaussian channel with arbitrary input is expressed as the solution of a fixed point equation. When the input signal is Gaussian distributed, the one-channel design problem can be solved as a singular value decomposition (SVD) problem.  More specially, the optimal channel matrix has its singular vectors allocated to create non-interfering subchannels and the  singular values chosen to solve a generalized waterfilling problem~\cite{Lamarca:2009}, \cite{Cover:Thomas:2005}. In \cite{Liu:Chong:Scharf:2012}, a greedy adaptive approach is considered to design a channel matrix row by row to maximize information gain.

In Fig.~\ref{fig:linsystem}, if both $\bm\theta$ and $\bm\phi$ are of interest, the two-channel system can be expressed as a one-channel system with a block-diagonal channel matrix. However, due to the ``nuisance'' signal $\bm\phi$,  our two-channel system design problem is fundamentally more difficult than the one-channel system design. In this paper, we fix the primary channel and design the secondary channel matrix $\bm G$ that maximizes the information gain brought by adding the secondary channel, subject to the total power constraint $\operatorname{tr}(\bm G\bm G^T)\leq P$ with $P$ a pre-determined constant. We call this a \textit{one-channel design problem in a two-channel system}.   Analytical solutions are derived for some special cases. In general, this is not a convex problem. Moreover, this problem cannot be formulated as an SVD problem, in contrast to the one-channel system design. Here, we propose two gradient-based algorithms, one extrinsic and the other intrinsic, to approximate the optimal channel matrix. The extrinsic algorithm is a gradient-ascent algorithm with projection to the constrained space \cite{Bertsekas:1982}. The intrinsic algorithm, a gradient-ascent algorithm of manifold, exploits the geometry that codes for the total power constraint by vectorizing the channel matrix, \cite{Absil:Mahony:Sepulchre:2008}, \cite{Edelman:Arias:Smith:1998}-\cite{Gabay:1982}.

 The rest of the paper is organized as follows. We formulate the channelization problem in Section~\ref{Sec:Formulation} and point out the challenges for design in a two-channel system.   In Section~\ref{sec:methodology}, we give an analytical solution when the conditional covariance of $\bm\phi$ given $\bm\theta$ is the identity matrix. In Section~\ref{Sec:IterAlgo}, we propose two numerical algorithms, the extrinsic and intrinsic gradient searches, to approximate the optimal channel matrix for general cases.   A simulation study is presented to illustrate the performance of the proposed algorithms in Section~\ref{sec:casestudy}. In Section~\ref{Sec:DimDisc}, we discuss the choice of number of measurements for the secondary channel. Section~\ref{Sec:Conclu} concludes the paper.

\textit{Notation}: The set of length $m$ real vectors is denoted by $\mathbb{R}^m$ and the set of $m \times n$ real matrices is denoted $\mathbb{R}^{m \times n}$. Bold upper case letters denote matrices, bold lower case letters denote column vectors, and italics denote scalars. The scalar $x_i$ denotes the $i$th element of vector $\bm x$,  and $\bm X_{i,j}$ denotes the element of $\bm X$ at row $i$ and column $j$.  The diagonal matrix with diagonal elements $\bm x$ is denoted as $\operatorname{Diag}(\bm x)$. The $n \times n$ identity matrix is denoted by $\bm I_n$. The transpose, inverse, trace and determinant of a matrix are denoted by $(\cdot)^T$, $(\cdot)^{-1}$, $\operatorname{tr}(\cdot)$ and $\det(\cdot)$, respectively.

A covariance matrix is denoted by  bold upper case $\bm Q$ with specified subscripts: $\bm Q_{\bm z\bm z}$ denotes the covariance matrix of a random vector $\bm z$; $\bm Q_{\bm z_1\bm z_2}$ is the cross-covariance matrix between $\bm z_1$ and $\bm z_2$; $\bm Q_{\bm z_1\bm z_1|\bm z_2}$ is the conditional covariance matrix of $\bm z_1$ given $\bm z_2$.

\section{Overview}\label{Sec:Formulation}

\subsection{Problem Statement}
The two channels of the system described in~\eqref{eqn:linsys} have input signals $\bm\theta\in\mathbb{R}^{p}$ and $\bm\phi\in\mathbb{R}^{q}$, respectively. The signal $\bm\theta$ is of key interest and $\bm\phi$ is a secondary signal that is jointly distributed with $\bm\theta$.  The first channel $\bm x\in\mathbb{R}^s$ is a direct measurement of $\bm\theta$, while the secondary channel $\bm y\in\mathbb{R}^{t}$ is an indirect measurement of $\bm\theta$ through $\bm\phi$. Both $\bm x$ and~$\bm y$ contain information about $\bm\theta$, and one can expect that fusing  measurements from both channels would provide a better estimate than using a single measurement.   The data fusion problem has been widely studied in various areas including sensor networks, image processing, etc. While much of the literature focuses  on the methodology of fusion or data integration, we are interested in designing the measurement system. More specifically, our interest is to design the channel matrix $\bm G$, with the first channel fixed, such that the rate at which $\bm x$ and $\bm y$ bring information about $\bm\theta$ is maximized.

We make the following assumptions:
\begin{itemize}
\item [$a1)$]  The signals $\bm\theta\in\mathbb{R}^p$ and $\bm\phi\in\mathbb{R}^q$ are jointly Gaussian distributed as
$$
 \left(\begin{array}{c} \bm\theta \\\bm\phi \end{array}\right)\sim N\left(\left(\begin{array}{c} \bm\mu_{\bm\theta} \\\bm \mu_{\bm\phi} \end{array}\right),\left(\begin{array}{cc} \bm Q_{\bm\theta\bm\theta} & \bm Q_{\bm\theta\bm\phi} \\ \bm Q_{\bm\phi\bm\theta}& \bm Q_{\bm\phi\bm\phi}\end{array}\right)\right )
$$
with known $\bm Q_{\bm\theta\bm\theta}$, $ \bm Q_{\bm\theta\bm\phi}$, $\bm Q_{\bm\phi\bm\theta}$ and $\bm Q_{\bm\phi\bm\phi}$.
\item [$a2)$] The noises $\bm{u}\in \mathbb{R}^s$ and $\bm{v}\in \mathbb{R}^t$ are Gaussian distributed with mean zero and known covariance matrices $\bm Q_{\bm u \bm u}$ and $\bm Q_{\bm v\bm v}$, respectively.
\item [$a3)$]The noises $\bm u$ and $\bm v$ are mutually independent, and independent of $(\bm\theta,\bm\phi)$.
\end{itemize}
Based on all the assumptions, the mutual information between $\bm\theta$ and $\bm x$ is
$$
I(\bm\theta;\bm x)=\frac{1}{2}\log\det(\bm Q_{\bm\theta\bm\theta})-\frac{1}{2}\log\det(\bm Q_{\bm\theta\bm\theta|\bm x}),
$$
where $\bm Q_{\bm\theta\bm\theta|\bm x}=(\bm Q_{\bm\theta\bm\theta}^{-1}+\bm F^T\bm Q_{\bm u\bm u}^{-1}\bm F)^{-1}$ is the conditional covariance of $\bm\theta$ given $\bm x$.
The mutual information between $\bm \theta$ and $\bm x,\bm y$ is
$$
I(\bm\theta;\bm x, \bm y)=\frac{1}{2}\log\det(\bm Q_{\bm\theta\bm\theta})-\frac{1}{2}\log\det(\bm Q_{\bm\theta\bm\theta|\bm x, \bm y}),
$$
where $\bm Q_{\bm\theta\bm\theta|\bm x,\bm y}=[\bm Q_{\bm\theta\bm\theta|\bm x}^{-1}+\bm M^T\bm G^T(\bm G\bm Q_{\bm\phi\bm\phi|\bm\theta}\bm G^T+\bm Q_{\bm v\bm v})^{-1}\bm G\bm M]^{-1}$ with $\bm M=\bm Q_{\bm\phi\bm\theta}\bm Q_{\bm\theta\bm\theta}^{-1}$ and $\bm Q_{\bm\phi\bm\phi|\bm\theta}=\bm Q_{\bm\phi\bm\phi}-\bm Q_{\bm\phi\bm\theta}\bm Q_{\bm\theta\bm\theta}^{-1}\bm Q_{\bm\theta\bm\phi}$. Note that $\bm M\bm\theta$ would be the MMSE estimator of $\bm\phi$ from $\bm\theta$, and $\bm Q_{\bm\phi\bm\phi|\bm\theta}$ would be its error covariance, if $\bm\theta$ could be measured.

The \textit{information gain} is the extra information about $\bm\theta$ brought by $\bm y$, defined as
\begin{align*}
D(\bm G):=I(\bm \theta;\bm x, \bm y)-I(\bm\theta; \bm x)=\frac{1}{2}\log\det\bm Q_{\bm\theta\bm\theta|\bm x}-\frac{1}{2}\log\det\bm Q_{\bm\theta\bm\theta|\bm x, \bm y}.
\end{align*}
By plugging in $\bm Q_{\bm\theta\bm\theta|\bm x}$ and $\bm Q_{\bm\theta\bm\theta|\bm x, \bm y}$, $D(\bm G)$ can be written as
\begin{eqnarray}\label{Eqn:infogain}
D(\bm G)=\frac{1}{2}\log\det[\bm I_p+\bm M^T\bm G^T(\bm G\bm Q_{\bm\phi\bm\phi|\bm\theta}\bm G^T+\bm Q_{\bm v\bm v})^{-1}\bm G\bm M\bm Q_{\bm\theta\bm\theta|\bm x}]
\end{eqnarray}

The function $D(\bm G)$ is bounded and nonnegative. In fact, one can show that $D(\bm G)\leq I(\bm\theta;\bm x,\bm\phi)-I(\bm\theta;\bm x)$, which means the maximum information gain the measurement $\bm y$ can bring is no greater than that could be brought by $\bm\phi$. We further notice that, for any~$\bm G$, $D(\lambda\bm G)$ is monotone increasing for $\lambda\geq0$. Therefore, without any constraint, maximization of the information gain in \eqref{Eqn:infogain} will lead to a trivial solution by letting the norm of $\bm G$ go to infinity.
 Here we maximize the information gain subject to the total power constraint $\operatorname{tr}(\bm G\bm G^T)\leq P$. This constraint bounds the total power of $\bm G\bm\phi$ since $\operatorname{tr}\mathbb{E}[\bm G\bm\phi\bm\phi^T\bm G^T]\leq P\operatorname{tr}\mathbb{E}[\bm\phi\bm\phi^T]$. In short, the problem of interest is
\begin{align}\label{Eqn:keyprob}
\bm G^\ast=\underset{\bm G\in\mathbb{R}^{t\times q}}{\arg\max}\text{ } D(\bm G) \text{ subject to }  \operatorname{tr}(\bm G\bm G^T)\leq P.
\end{align}

Problem~\eqref{Eqn:keyprob} is a one-channel design problem in a two-channel system. In general, the optimization problem cannot be reformulated as an SVD problem in contrast to a one-channel system. The difficulty arises due to the non-degenerate joint distribution of $\bm\theta$ and $\bm\phi$. However, when the conditional covariance matrix $\bm Q_{\bm\phi\bm\phi|\bm\theta}$ is zero, i.e., the value of $\bm\phi$ is fixed given $\bm\theta$, the optimal channel matrix $\bm G$ can be solved from an SVD problem, as  in a one-channel system.

\subsection{An Insightful Discussion of the Information Gain}\label{sec:infogain}
\begin{figure}[htbp]
\centering
\includegraphics[totalheight=2in]{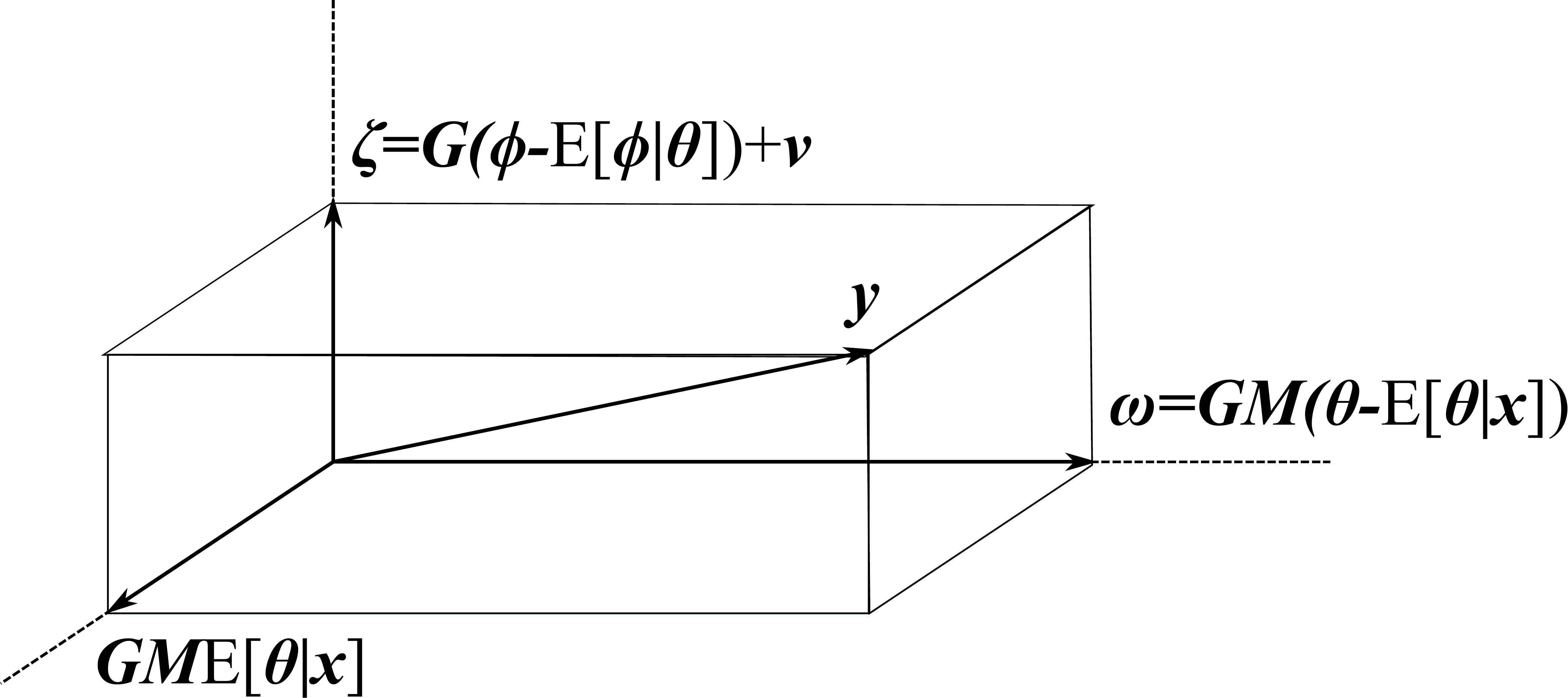}
\caption{Decomposition of the secondary channel.}
\label{fig:decomposition}
\end{figure}

To motivate our discussion, we decompose the secondary channel as follows:
\begin{eqnarray}\label{Eqn:ChannelDecom}
\bm y=\left(\bm G\bm M \mathbb{E}[\bm\theta|\bm x]\right)+\left(\bm G \bm M(\bm\theta-\mathbb{E}[\bm\theta|\bm x])\right)+\left(\bm G(\bm\phi-\mathbb{E}[\bm\phi|\bm\theta])+\bm v\right),
\end{eqnarray}
where $\bm M=\bm Q_{\bm\phi\bm\theta}\bm Q_{\bm\theta\bm\theta}^{-1}$ and  $\bm M\bm\theta=\mathbb{E}[\bm\phi|\bm\theta]$. It can be seen that the secondary channel $\bm y$ is decomposed into three independent components, which are illustrated in Fig.~\ref{fig:decomposition}. The first component $\bm G\bm M \mathbb{E}[\bm\theta|\bm x]$ is completely determined by the first channel $\bm x$ and does not contribute to the information gain brought by $\bm y$. The second component $\bm G \bm M(\bm\theta-\mathbb{E}[\bm\theta|\bm x])$, denoted by~$\bm\omega$, is (by orthogonality) independent of $\bm x$ and it carries the extra information in channel $\bm y$ about $\bm\theta$. The third component  $\bm G(\bm\phi-\mathbb{E}[\bm\phi|\bm\theta])+\bm v$, denoted by $\bm\zeta$, is independent of both $\bm x$ and $\bm\theta$, and it can be viewed as noise.

 Notice that the covariance matrices of $\bm\omega$ and $\bm\zeta$ are $\bm Q_{\bm\omega\bm\omega}=\bm G\bm M \bm Q_{\bm\theta\bm\theta|\bm x}\bm M^T\bm G^T$ and $\bm Q_{\bm\zeta\bm\zeta}=\bm G\bm Q_{\bm\phi\bm\phi|\bm\theta}\bm G^T+\bm Q_{\bm v \bm v}$, respectively. By the cyclic property of determinants, $\det(\bm I_m+\bm A\bm B)=\det(\bm I_n +\bm B\bm A)$ for any $\bm A\in\mathbb{R}^{m\times n}$ and $\bm B\in\mathbb{R}^{n\times m}$, the information gain of \eqref{Eqn:infogain} can be re-written~as
\begin{eqnarray}\label{Eqn:SNRMat}
D(\bm G)=\frac{1}{2}\log\det [\bm I+\bm Q_{\bm\zeta\bm\zeta}^{-1/2}\bm Q_{\bm\omega\bm\omega}\bm Q_{\bm\zeta\bm\zeta}^{-1/2}].
\end{eqnarray}
By viewing $\bm\omega$ as a signal and $\bm\zeta$ as a noise, $\bm Q_{\bm\zeta\bm\zeta}^{-1/2}\bm Q_{\bm\omega\bm\omega}\bm Q_{\bm\zeta\bm\zeta}^{-1/2}$ is a generalized signal-to-noise ratio matrix. Maximizing \eqref{Eqn:SNRMat} essentially balances the tradeoff between the noise covariance and the signal covariance. As illustrated in Fig.~\ref{fig:decomposition}, a good channel design will favor a long parallelepiped with short height. The difficulty of designing the channel matrix $\bm G$ arises because $\bm G$ shapes both $\bm Q_{\bm\omega\bm\omega}$ and $\bm Q_{\bm\zeta\bm\zeta}$. When the secondary channel  has a single output, the channel matrix $\bm G$ is a row vector and $\bm Q_{\bm\zeta\bm\zeta}, \bm Q_{\bm\omega\bm\omega}$ are scalars. In this case, the optimal $\bm G$ equivalently maximizes a generalized Rayleigh quotient and the analytical solution can be derived by an eigendecomposition. For a general channel with multiple outputs, designing the matrix $\bm G$ is fundamentally more difficult than the single output case.  In Section~\ref{sec:methodology}, we obtain a closed-form expression for the optimal $\bm G$ in a special case.

\section{Analytical Solution }\label{sec:methodology}

 Suppose that the conditional covariance of $\bm\phi$ given $\bm\theta$ is identity, i.e., $\bm Q_{\bm\phi\bm\phi|\bm\theta}=\sigma^2_{\bm\phi|\bm\theta}\bm I_q$. For example, $\bm\phi=\bm M\bm\theta+\bm\tau$ where $\bm\tau\sim N(\bm 0, \sigma^2_{\bm\phi|\bm\theta}\bm I_q)$. In this case, the noise $\bm\zeta$ in~\eqref{Eqn:SNRMat} has a relatively simple covariance $\bm Q_{\bm\zeta\bm\zeta}=\bm G\bm G^T+\bm Q_{\bm v \bm v}$ and the signal $\bm\omega$ has covariance $\bm Q_{\bm\omega\bm\omega}=\bm G\bm M \bm Q_{\bm\theta\bm\theta|\bm x}\bm M^T\bm G^T$. While $\bm G$ still affects both covariance matrices, we are able to find the balanced matrix $\bm G$ that maximizes  the information gain. Note that we focus on the case $t\leq q$, i.e., the dimension of measurement $\bm y$ is at most the dimension of $\bm\phi$.  When $t>q$, the optimization problem  can be reformulated and solved as a special case of $t=q$, which will be discussed in Section~\ref{Sec:DimDisc}.

Given the eigendecompositions $\bm Q_{\bm v\bm v}=\bm U_{\bm v}\bm\Sigma_{\bm v}\bm U_{\bm v}$ and $\bm M \bm Q_{\bm\theta\bm\theta|\bm x}\bm M^T=\bm U_{\bm\xi}\bm\Sigma_{\bm\xi}\bm U_{\bm\xi}^T$ where $\bm U_{\bm v}\in\mathbb{R}^{t\times t }$, $\bm U_{\bm\xi}\in\mathbb{R}^{q\times q }$ are orthogonal matrices, and $\bm\Sigma_{\bm v}\in\mathbb{R}^{t\times t }$ and $\bm\Sigma_{\bm\xi}\in\mathbb{R}^{q\times q }$ are  diagonal matrices with diagonal elements $0<\sigma_{\bm v, 1}^2\leq\ldots\leq\sigma_{\bm v,m}^2$ and $\sigma_{\bm\xi, 1}^2\geq\ldots\geq\sigma_{\bm\xi,q}^2\geq 0$, respectively. Because the matrices $\bm U_{\bm\xi}$ and $\bm U_{\bm v}$ are invertible, for each $\bm G\in\mathbb{R}^{t\times q}$,  there is a unique matrix  $\bm\Phi\in\mathbb{R}^{t \times q}$ such that
\begin{align}\label{Eqn:GPhi}
\bm G=\bm U_{\bm v}\bm\Phi\bm U_{\bm\xi}^T
\end{align}
Then, the information gain $D(\bm G)$ in~\eqref{Eqn:infogain} can be written as
$$
D(\bm\Phi)=\frac{1}{2}\log\det[\bm I+\bm\Phi^T(\sigma^2_{\bm\phi|\bm\theta}\bm\Phi\bm\Phi^T+\bm\Sigma_{\bm v})^{-1}\bm\Phi\bm\Sigma_{\bm\xi}]
$$
  Moreover, the total power constraint is $\operatorname{tr}(\bm\Phi\bm\Phi^T)\leq P$ since $\operatorname{tr}(\bm G\bm G^T)=\operatorname{tr}(\bm\Phi\bm\Phi^T)$.  For the given eigendecompositions, the matrices $\bm U_{\bm v}$ and $\bm U_{\bm\xi}$ are fixed. Therefore, the information gain can be maximized with respect to $\bm\Phi$ and the optimal channel matrix $\bm G$ is returned by~\eqref{Eqn:GPhi}. WOLG we assume $\sigma^2_{\bm\phi|\bm\theta}=1$. The solution for general $\sigma^2_{\bm\phi|\bm\theta}$ is just different by a scaling factor.  We give in Lemma~\ref{lemma:phi} an important feature of any possible maximizer $\bm\Phi$.
\begin{lemma}\label{lemma:phi}
 Suppose that $\bm Q_{\bm v\bm v}$ has distinct eigenvalues, i.e., $0<\sigma_{\bm v, 1}^2<\ldots<\sigma_{\bm v,m}^2$, and $\bm M \bm Q_{\bm\theta\bm\theta|\bm x}\bm M^T$ has distinct nonzero eigenvalues, i.e., $\sigma_{\bm\xi,1}^2>\ldots>\sigma_{\bm\xi,\rho}^2>0$ where $\rho\leq t$ is the rank of $\bm\Sigma_{\bm\xi}$. Then  $\bm\Phi$  contains at most one nonzero entry in each row and column and all  the nonzero entries are located at the first $\rho$ columns.
\end{lemma}
Proof: See Appendix~\ref{Appen:LemmaPhi}.

Lemma~\ref{lemma:phi} restricts the optimal matrix $\bm\Phi$ within a class of matrices with a special structure. That is, $\bm\Phi$  has at most one nonzero entry in each row and column. Searching within this class, we are able to obtain the closed form expression for the optimal matrix~$\bm\Phi$. The corresponding optimal channel matrix $\bm G$ is given in Theorem~\ref{thm:Gstar}.

\begin{Theorem}\label{thm:Gstar}
 Suppose that $\bm Q_{\bm v\bm v}$ and $\bm M \bm Q_{\bm\theta\bm\theta|\bm x}\bm M^T$ have distinct nonzero eigenvalues. Then the optimal secondary channel matrix $\bm G^\ast$  solving problem~\eqref{Eqn:keyprob} is
\begin{align}\label{Eqn:Wdet}
\bm G^\ast=\bm U_{\bm v}\bm\Lambda^\ast\bm U_{\bm\xi}^T.
\end{align}
Here $\bm\Lambda^\ast\in\mathbb{R}^{t\times q}$
is a diagonal matrix with diagonal elements $\lambda_{11}^\ast,\ldots,\lambda_{tt}^\ast$ such that
\begin{align}\label{Eqn:phi0}
\lambda_{ii}^{\ast2}=
\left\{\begin{array}{ll}
\frac{\sigma^2_{\bm v, i }}{2(1+\sigma^2_{\bm\xi, i })}\left(-(2+\sigma^2_{\bm\xi, i })+\sqrt{\sigma^4_{\bm\xi, i }+\frac{4(1+\sigma^2_{\bm\xi, i })\sigma^2_{\bm\xi, i }}{ 2\mu \sigma^2_{\bm v, i } }}\right) & i=1,\ldots,\kappa\\
0 &i=\kappa+1,\ldots,t
\end{array}
\right.
\end{align}
where
$\kappa$ is the maximum integer between $1$ and $\operatorname{rank}(\bm\Sigma_{\bm\xi})$ such that $\lambda_{ii}^{\ast2}>0$ for $i=1,\ldots,\kappa$. The value of $\mu$ is non-negative and uniquely solves $\sum_{i=1}^\kappa\lambda_{ii}^{\ast2}=P$.
\end{Theorem}
Proof: See Appendix~\ref{Appen:Gstar}.

 Notice that although Theorem~\ref{thm:Gstar} requires that $\bm Q_{\bm v\bm v}$ and $\bm M \bm Q_{\bm\theta\bm\theta|\bm x}\bm M^T$ have distinct eigenvalues, the result can be extended to general cases because the solution in~\eqref{Eqn:phi0} is a continuous function of the eigenvalues of $\bm Q_{\bm v\bm v}$ and $\bm M \bm Q_{\bm\theta\bm\theta|\bm x}\bm M^T$.

 Theorem~\ref{thm:Gstar} factors the optimal channel matrix $\bm G^\ast$ into the product of three matrices. The first matrix~$\bm U_{\bm\xi}^T$ rotates the signal $\bm\phi$. Given $\bm Q_{\bm\phi\bm\phi|\bm\theta}=\sigma^2_{\bm\phi|\bm\theta}\bm I_q$, the conditional covariance of $\bm\phi$ given $\bm x$ is $\bm M\bm Q_{\bm\theta\bm\theta|\bm x}\bm M^T+\sigma^2_{\bm\phi|\bm\theta}\bm I_q$, which is diagonalized by $\bm U_{\bm\xi}^T$. Therefore,   the components of the rotated signal $\bm U_{\bm\xi}^T\bm\phi$ are conditionally independent  given $\bm x$. The second matrix $\bm\Lambda^\ast\in\mathbb{R}^{t\times q}$ is a diagonal matrix that extracts the first $t$ components of $\bm U_{\bm\xi}^T\bm\phi$ and distributes power across the $t$ subchannels optimally. The third matrix $\bm U_{\bm v}$ rotates the scaled components into the sub-dominant invariant subspace of the noise covariance $\bm Q_{\bm v\bm v}$.

The power allocation policy, given by the diagonal elements of $\bm\Lambda^\ast$, can be interpreted as a \textit{mercury/waterfilling} algorithm, which is a three-step procedure that has been introduced in \cite{Lozano:Tulino:Verdu:2006}:
\begin{enumerate}
\item For the $i$th vessel,  fill in the solid base with
height $2\sigma^2_{\bm v, i }/\sigma^2_{\bm\xi, i }$, where $\sigma^2_{\bm v, i }$ is a noise variance component in the $\bm y$ channel, and $\sigma^2_{\bm\xi, i }$ yields a variance component of $\bm\phi$ given $\bm x$.
\item Compute $\mu$. For the vessels with base height less than $1/\mu$, fill in mercury in the vessel until the height reaches
\[\max\left\{\frac{1}{\mu}-\lambda_{ii}^{\ast 2},\frac{2\sigma^2_{\bm v, i }}{\sigma^2_{\bm\xi, i }}\right\}\].
\item Pour water into all the vessels  until the height of each vessel reaches $1/\mu$.
\end{enumerate}

 The height of the solid base, $2\sigma^2_{\bm v, i }/\sigma^2_{\bm\xi, i }$, is half of the variance of the $i$th noise weighted by the variance components of $\bm M\bm Q_{\bm\theta\bm\theta|\bm x}\bm M^T$.  A higher solid base means a less informative channel with high channel noise and weak correlation with $\bm\theta$. For any vessel with base height exceeding $1/\mu$, neither mercury nor water will be added, or equivalently, no power will be assigned to the corresponding subchannel. Note that the value of $\mu$ is computed by the constraint that the total volume of water equals $P$. The mercury stage balances the noise contained in $\bm\phi$ and the measurement noise contained in $\bm y$. Without adding mercury, the optimal power allocation will have variable water-plus-solid levels among different vessels. The mercury is added to regulate the water level for each vessel. Given the value of $\mu$, the information gain is maximized when the value of $\lambda_{ii}^{\ast2}$ equals the height of water in the corresponding vessel.

\begin{figure}[htbp]
\centering
\includegraphics[width=4in]{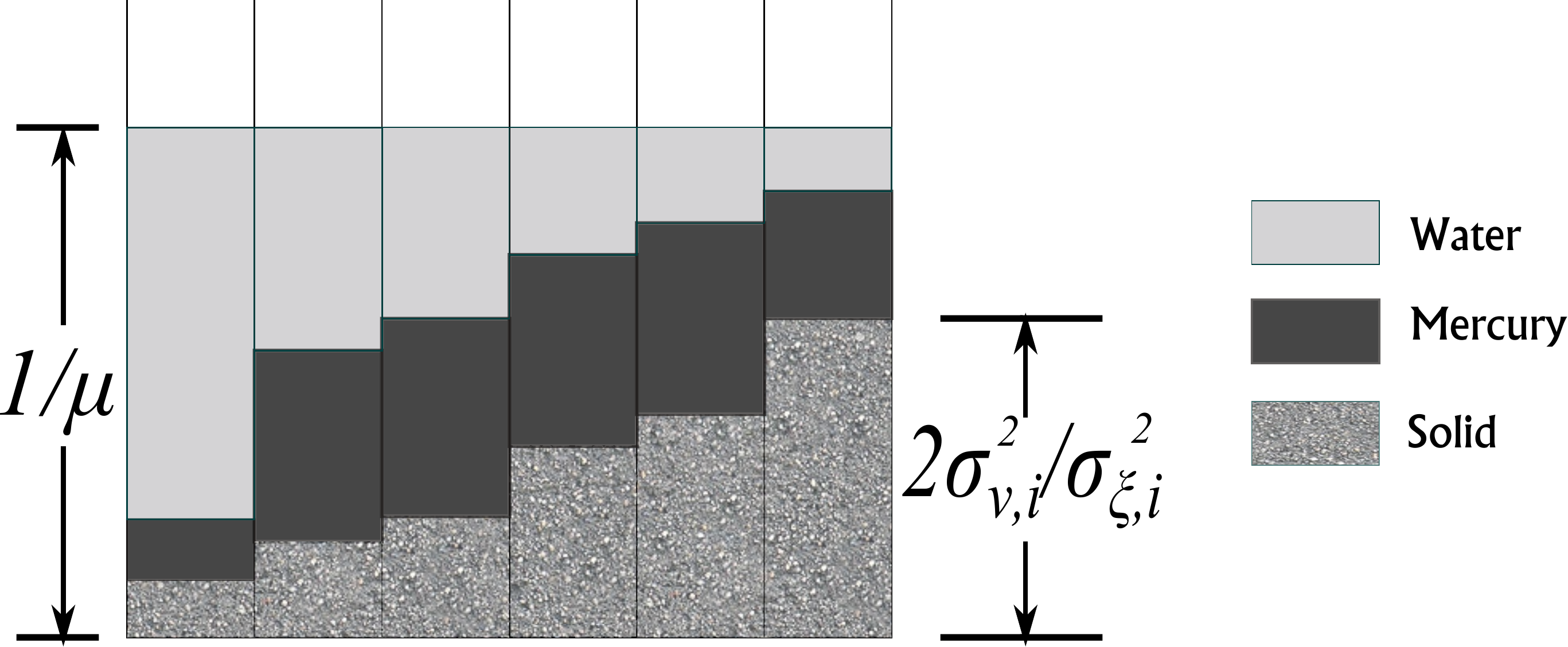}
\caption{Mercury/waterfilling. For each vessel, the water height above mercury gives the optimal power allocation to the corresponding subchannel.  The total volume of water equals $P$.}\label{fig:waterfilling}
\end{figure}

From the mercury/waterfilling procedure, it can be seen that the resulting optimal channel matrix~$\bm G^\ast$ may not be full-rank. We will see in Section~\ref{Sec:DimDisc} that a rank-reduced channel matrix can in some cases  give a dimension-reduced secondary channel that carries the same information gain as a full-dimensional channel, under the power constraint. To better illustrate the possible rank-reduced optimal channel, we consider the following simple example.

\begin{example} Consider a two-channel system in \eqref{eqn:linsys} with $p=q=s=t=5$. The primary channel matrix $\bm F\in\mathbb{R}^{5\times5}$ is set as $\frac{1}{\sqrt{5}}\bm I_5$. The covariance matrices $\bm Q_{\bm u\bm u}$, $\bm Q_{\bm v\bm v}$, and $\bm Q_{\bm\theta\bm\theta}$ are $\bm I_{5}$. %, and $\bm Q_{\bm\theta\bm\theta|\bm x}=\frac{5}{6}\bm I_5$.
We consider three scenarios. In each scenario, we choose $\bm Q_{\bm\phi\bm\phi}$ and $\bm Q_{\bm\phi\bm\theta}$ such that $\bm Q_{\bm\phi\bm\phi|\bm\theta}=\bm I_{5}$ and the eigenvalues of $\bm M\bm Q_{\bm\theta\bm\theta|\bm x}\bm M^T$ have various levels of spread. The corresponding $\bm G^\ast$ is given in Table~I.

In the first scenario, $\bm M\bm Q_{\bm\theta\bm\theta|\bm x}\bm M^T$ has constant eigenvalues and $\bm G^\ast$ has full-rank and equal singular values. In the second scenario, the eigenvalues of  $\bm M\bm Q_{\bm\theta\bm\theta|\bm x}\bm M^T$ have moderate spread and the corresponding $\bm G^\ast$ has rank $4$. In the third scenario, when the spread of eigenvalues of $\bm M\bm Q_{\bm\theta\bm\theta|\bm x}\bm M^T$ further increases, the rank of $\bm G^\ast$ is further reduced to $3$.
\begin{table}[htbp]
\begin{eqnarray*}
\begin{array}{ccc}
\hline
& \bm M\bm Q_{\bm\theta\bm\theta|\bm x}\bm M^T & \bm G^\ast\\
 \hline
 (1)&\frac{5}{6}\bm I_5&\frac{1}{\sqrt{5}}\bm I_5\\
 \hline
 (2)&\frac{5}{6}\operatorname{Diag}(25,16,9,4,1)&\operatorname{Diag}(0.316, 0.294, 0.249, 0.141,0)\\
 \hline
 (3)&\frac{5}{6}\operatorname{Diag}(81,64,49,4,1)&\operatorname{Diag}(0.338, 0.334, 0.328,0,0)\\
 \hline
 \end{array}
 \end{eqnarray*}
\caption{The optimal channel matrices $\bm G^\ast$ for three scenarios.}
 \end{table}
\end{example}

\begin{figure}[htbp]
\centering
\includegraphics[totalheight=1.5in]{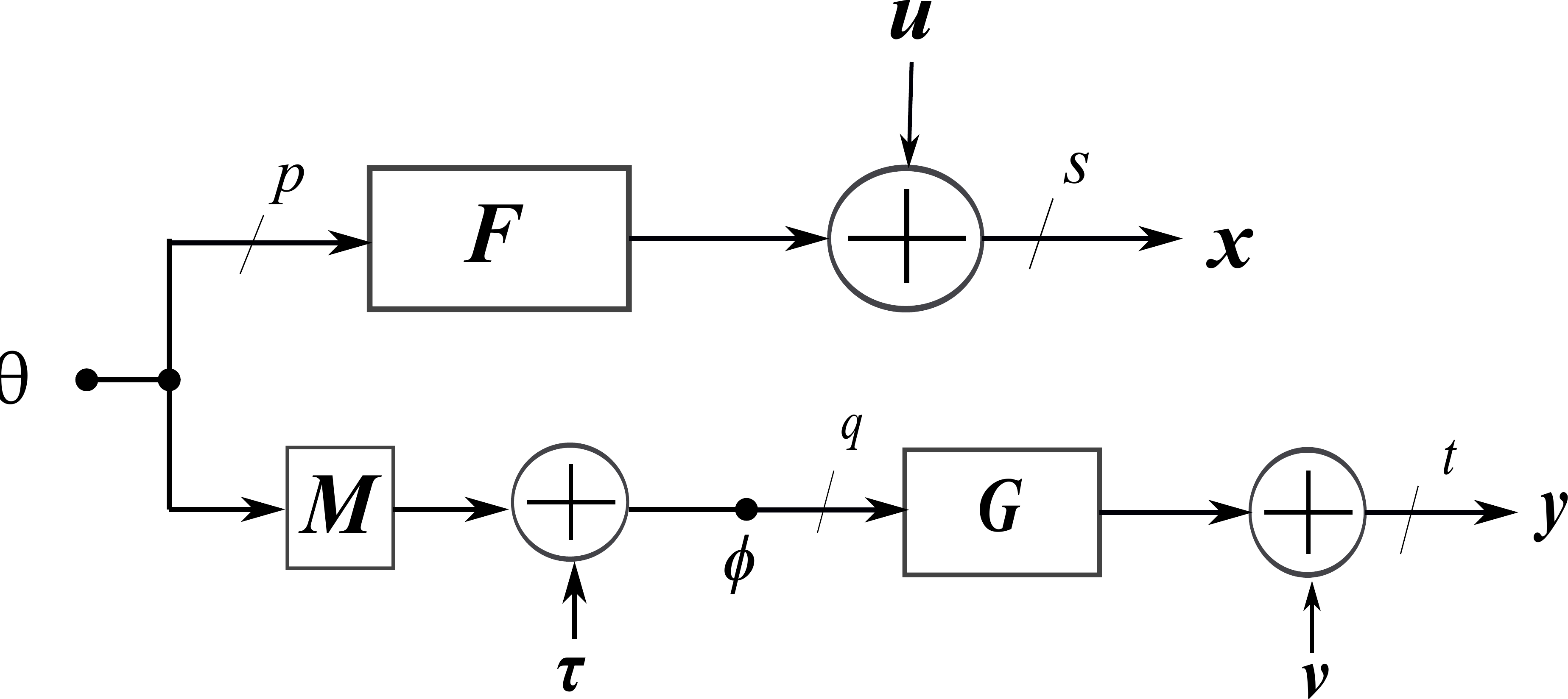}
\caption{An alternative representation for the two-channel system.}
\label{fig:linsys2}
\end{figure}

In \cite{Wang:Wang:Scharf:2013}, Wang et al.  generalized the problem of reduced-rank filtering and precoding/equalizing be designing the matrix $\bm G$ in the bottom channel of Fig.~\ref{fig:linsys2} so that $\bm y$ maximizes the differential rate at which $\bm y$ brings information about $\bm\theta$. The difference between that design and the fusion design of this paper is that there was no existing channel $\bm x$ to be fused with $\bm y$. Thus, the optimal channel matrix $\bm G$ needs to maximize the extra information $\bm y$ contains about $\bm\theta$, taking account of the primary channel $\bm x$.

\section{Numerical Algorithms}\label{Sec:IterAlgo}%

In general, the constrained optimization problem \eqref{Eqn:keyprob} is not a convex problem \cite{Boyd:Vandenberehe:2004} since the information gain $D(\bm G)$ is not concave \cite{Payaro:Palomar:2009}. Notice that for any  $\bm G$ with $\operatorname{tr}(\bm G\bm G^T)<P$, there exists $\widetilde{\bm G}=\frac{\sqrt{P}}{\|\bm G\|}\bm G$ such that $\operatorname{tr}(\widetilde{\bm G}\widetilde{\bm G}^T)=P$ and $D(\widetilde{\bm G})\geq D(\bm G)$. Therefore, it is sufficient to maximize the information gain on the boundary
$\operatorname{tr}(\bm G\bm G^T)=P$. This fact motivates two  gradient-based search algorithms, one extrinsic and the other intrinsic, to approximate the optimal channel matrix. Both algorithms are very general and applicable for arbitrary covariance between $\bm\theta$ and $\bm\phi$. In the extrinsic gradient search, the gradient is computed by treating the matrix $\bm G$ as a point in the Euclidean space~$\mathbb{R}^{t\times q}$. In the intrinsic gradient search, we consider $\bm G$ as a point on the unit sphere $S^{tq-1}$, which is a submanifold of~$\mathbb{R}^{tq}$. The intrinsic gradient is computed by taking the geometry of the manifold $S^{tq-1}$ into consideration.  WLOG we assume $P=1$.

\subsection{Extrinsic Gradient Search Algorithm}\label{sec:ExAlgo}

 Let $\nabla_{\bm G}D$ be the gradient of the information  gain w.r.t $\bm G$, and it can be written as
\begin{eqnarray}\label{Eqn:GradI}
\nabla_{\bm G} D=\bm Q_{\bm v\bm v}^{-1}\bm G[(\bm Q_{\bm\phi\bm\phi|\bm\theta}^{-1}+\bm G^T\bm Q_{\bm v\bm v}^{-1}\bm G)-\bm B]^{-1}\bm B(\bm Q_{\bm\phi\bm\phi|\bm\theta}^{-1}+\bm G^T\bm Q_{\bm v\bm v}^{-1}\bm G)^{-1},
\end{eqnarray}
where $\bm B=\bm Q_{\bm\phi\bm\phi|\bm\theta}^{-1}\bm M\bm Q_{\bm\theta\bm\theta|\bm x}\bm M^T(\bm I_q+ \bm Q_{\bm\phi\bm\phi|\bm\theta}^{-1}\bm M\bm Q_{\bm\theta\bm\theta|\bm x}\bm M^T)^{-1}\bm Q_{\bm\phi\bm\phi|\bm\theta}^{-1}$. See Appendix \ref{Appen:Gradient} for details. The gradient $\nabla_{\bm G}D$ points in the direction of greatest increase of the function $D$ in the neighborhood of $\bm G$. However, when moving along this direction, the constraint $\operatorname{tr}(\bm G\bm G^T)=1$ may be violated. To circumvent this problem, we normalize the updated $\bm G$ at each iteration to meet the unit norm constraint. The table below outlines the proposed extrinsic gradient search algorithm.
\begin{center}
\begin{tabular}{l}
\hline
Algorithm: Extrinsic Gradient Search\\
\hline
{\bf Input:} Initial $\bm G_0\in\mathbb{R}^{t \times q}, \operatorname{tr}(\bm G_0\bm G_0^T)=1$.\\
{\bf Output:} Sequence of iterates $\{\bm G_k\}$.\\
 {\bf for} $k=0,1,2,\ldots$ {\bf do} \\
Select $\bm G_{k+1}=a_k(\bm G_k+\delta_k \nabla_{\bm G_k}D)$ where $a_k=\frac{1}{\|\bm G_k+\delta_k \nabla_{\bm G_k}D\|}$ is  a \\ normalization constant such that
 $\operatorname{tr}({\bm G_{k+1}\bm G_{k+1}^T})=1$,  $\delta_k$ is a small step size.\\
{\bf end for} \\
\hline
\end{tabular}
\end{center}

In this extrinsic algorithm, the gradient of the information gain is computed on the unconstrained Euclidean space $\mathbb{R}^{t \times q }$. 
Note that $\bm G_k+\delta_k \nabla_{\bm G_k}D$ is the unconstrained update when maximizing $D$. The normalized update $\bm G_{k+1}=a_k(\bm G_k+\delta_k \nabla_{\bm G_k}D)$ is a projection of $\bm G_k+\delta_k \nabla_{\bm G_k}D$ onto the set of all $\bm G\in\mathbb{R}^{t \times q}$ with unit Frobenius norm.

We call it an extrinsic gradient search in contrast to the intrinsic gradient search algorithm, in which the information gain is considered as a function on the manifold $S^{tq-1}$.
\subsection{Intrinsic Gradient Search Algorithm}\label{sec:IntAlgo}

Let $\bm g$ be the vectorization of matrix $\bm G$, denoted $\bm g=\operatorname{vec}(\bm G)$. That is,
$$\bm g=[\bm G_{1,1},\ldots,\bm G_{1,q},\bm G_{2,1},\ldots,\bm G_{2,q},\ldots,\bm G_{t,1},\ldots,\bm G_{t,q}]^T.$$
This vectorization operation is a one-to-one and onto mapping from $\mathbb{R}^{t\times q}$ to $\mathbb{R}^{tq}$. Thus, for any $\bm g\in\mathbb{R}^{tq}$, there exists a unique matrix $\bm G\in\mathbb{R}^{t \times q}$ such that $\operatorname{vec}(\bm G)=\bm g$. Under the power constraint $\operatorname{tr}(\bm G\bm G^T)=1$, the corresponding vectorization $\bm g$ lies on the unit sphere $S^{tq-1}=\{\bm g \in\mathbb{R}^{tq}: \sum_{i=1}^{tq} g_i^2=1\}$. Therefore, the constrained optimization problem~\eqref{Eqn:keyprob} is an optimization on the manifold $S^{tq-1}$.
Note that $S^{tq-1}$ is an embedded submanifold of $\mathbb{R}^{tq}$, and its geometry has been studied in \cite{Absil:Mahony:Sepulchre:2008}-\cite{Lee:2000}.

\begin{figure}[htbp]
\centering
\includegraphics[totalheight=2.5in]{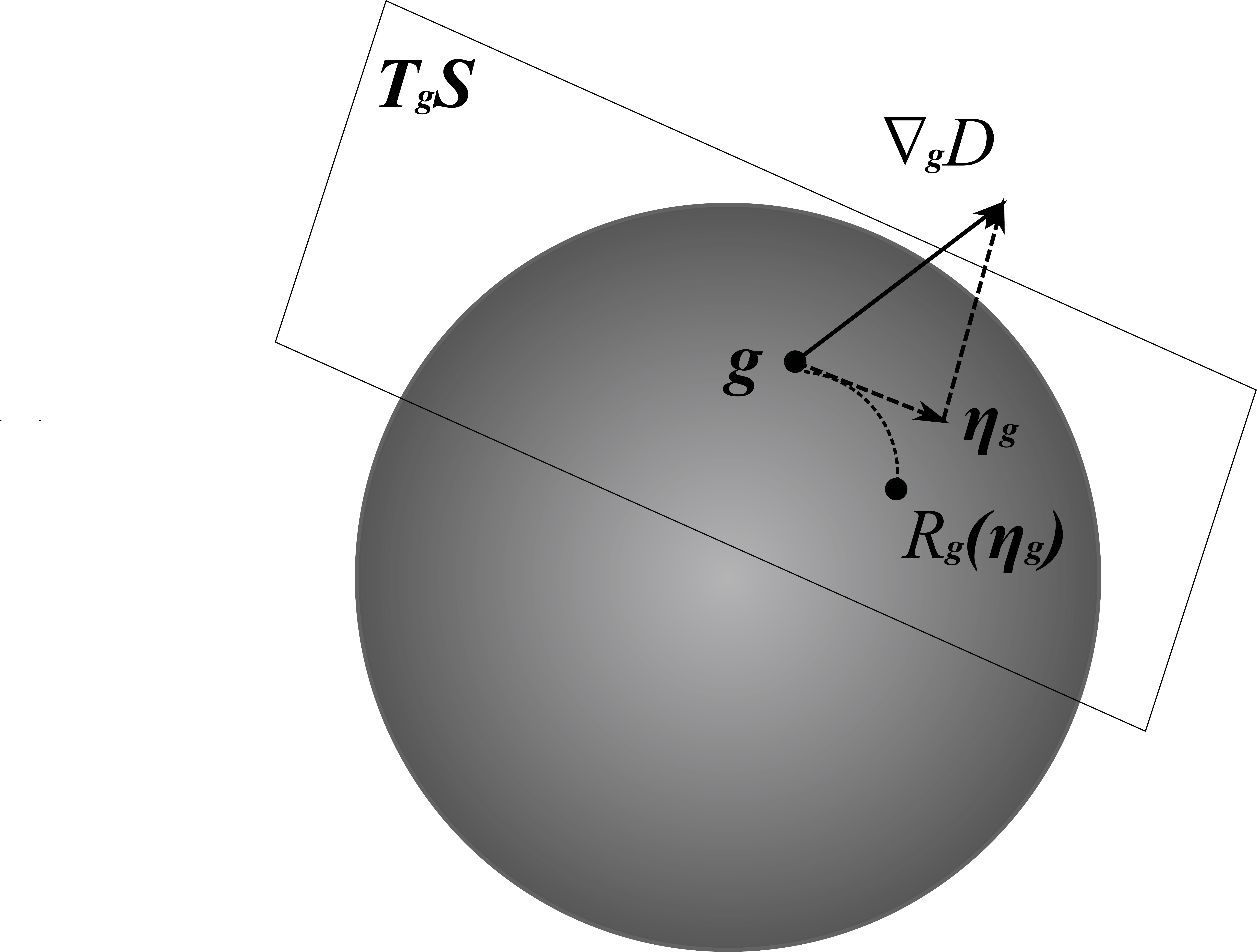}
\caption{Projection of the Euclidean gradient to the tangent plane of unit sphere.}
\label{fig:tangent}
\end{figure}

The following algorithm encodes the intrinsic gradient search, which approximates a maximizer of the information gain on the manifold~$S^{tq-1}$. A graphical illustration is depicted in Fig.~\ref{fig:tangent}.
\begin{center}
\begin{tabular}{l}
\hline
Algorithm: Intrinsic Gradient Search\\
\hline
{\bf Input:} Initial $\bm g_0\in S^{tq-1}$\\
{\bf Output:} Sequence of iterates $\{\bm g_k\}$.\\
 {\bf for} $k=0,1,2,\ldots$ {\bf do} \\
Select $\bm g_{k+1}=R_{\bm g_k}(\delta_k\bm\eta_{\bm g_k})$ where $\bm\eta_{\bm g_k}=(\bm I_{tq}-\bm g_k \bm g_k^T)\nabla_{\bm g_k}D$ is the intrinsic \\ gradient and
 $\delta_k$ is the step size.\\
{\bf end for} \\
\hline
\end{tabular}
\end{center}
For any $\bm g\in S^{tq-1}$, the tangent plane $T_{\bm g}S^{tq-1}$ is the subspace orthogonal to $\bm g$. The intrinsic gradient, denoted by $\bm\eta_{\bm g}$, is the Euclidean gradient $\nabla_{\bm g} D$ projected onto the tangent plane $T_{\bm g}S^{tq-1}$. The function $R_{\bm g}$ is a mapping from the tangent plane  $T_{\bm g}S^{tq-1}$ to the manifold $S^{tq-1}$ with \begin{eqnarray}\label{eqn:retraction}
R_{\bm g}(\bm\eta_{\bm g})=\bm g \cos(\|\bm\eta_{\bm g}\|)+ \frac{\bm\eta_{\bm g}}{\|\bm\eta_{\bm g}\|}\sin(\|\bm\eta_{\bm g}\|)
\end{eqnarray}
for any tangent vector $\bm\eta_{\bm g}\in T_{\bm g}S^{tq-1}$. For $\delta\geq0$, $R_{\bm g}(\delta\bm\eta_{\bm g})$ is a curve on the manifold $S^{tq-1}$ starting from $\bm g$. This curve generalizes the idea of straight line in Euclidean space on the manifold $S^{tq-1}$ along the direction $\bm\eta_{\bm g}$.
Given $\bm g_k$, $R_{\bm g}(\delta\bm\eta_{\bm g_k})$ is a periodic function of $\tau$ with period ${2\pi}/\|\bm\eta_{\bm g_k}\|$, thus the step size $\delta_k$ can be chosen within the interval $\delta\in[0,{2\pi}/\|\bm\eta_{\bm g_k}\|)$ to maximize the information gain $D(R_{\bm g}(\delta\bm\eta_{\bm g_k}))$. By the choice of $\delta_k$, the information gain is non-decreasing, i.e., $D(\bm g_{k+1})\geq D(\bm g_k)$ for each~$k$.

\subsection{A Numerical Study}\label{sec:casestudy}

Consider a two-channel system in \eqref{eqn:linsys} with $p=q=4$ and $s=t=3$. The input signals $\bm\theta$ and~$\bm\phi$ are characterized recursively as
$$
\phi_i=\sum_{j=1}^i\rho^{i-j+1}\theta_j+\tau_i,
$$
where $\tau_1,\ldots, \tau_4$ are i.i.d. Gaussian random variables with mean $0$ and variance $1$,  and  the value of $\rho$ is to be specified.
The covariance matrices for the signal $\bm\theta$ and the noises $\bm u, \bm v$ are proportional to the identity matrix with variance $2$, $1$, $0.1$, respectively.
The first channel matrix $\bm F\in\mathbb{R}^{3 \times 4}$ is a diagonal matrix with $1$ on the diagonal.  The initial channel matrix $\bm G_0\in\mathbb{R}^{3 \times 4}$ is randomly generated with unit norm.  For the intrinsic algorithm, the initial value is $\bm g_0=\operatorname{vec}(\bm G_0)$.

\begin{figure}[htbp]
\centering
\includegraphics[totalheight=2.3in]{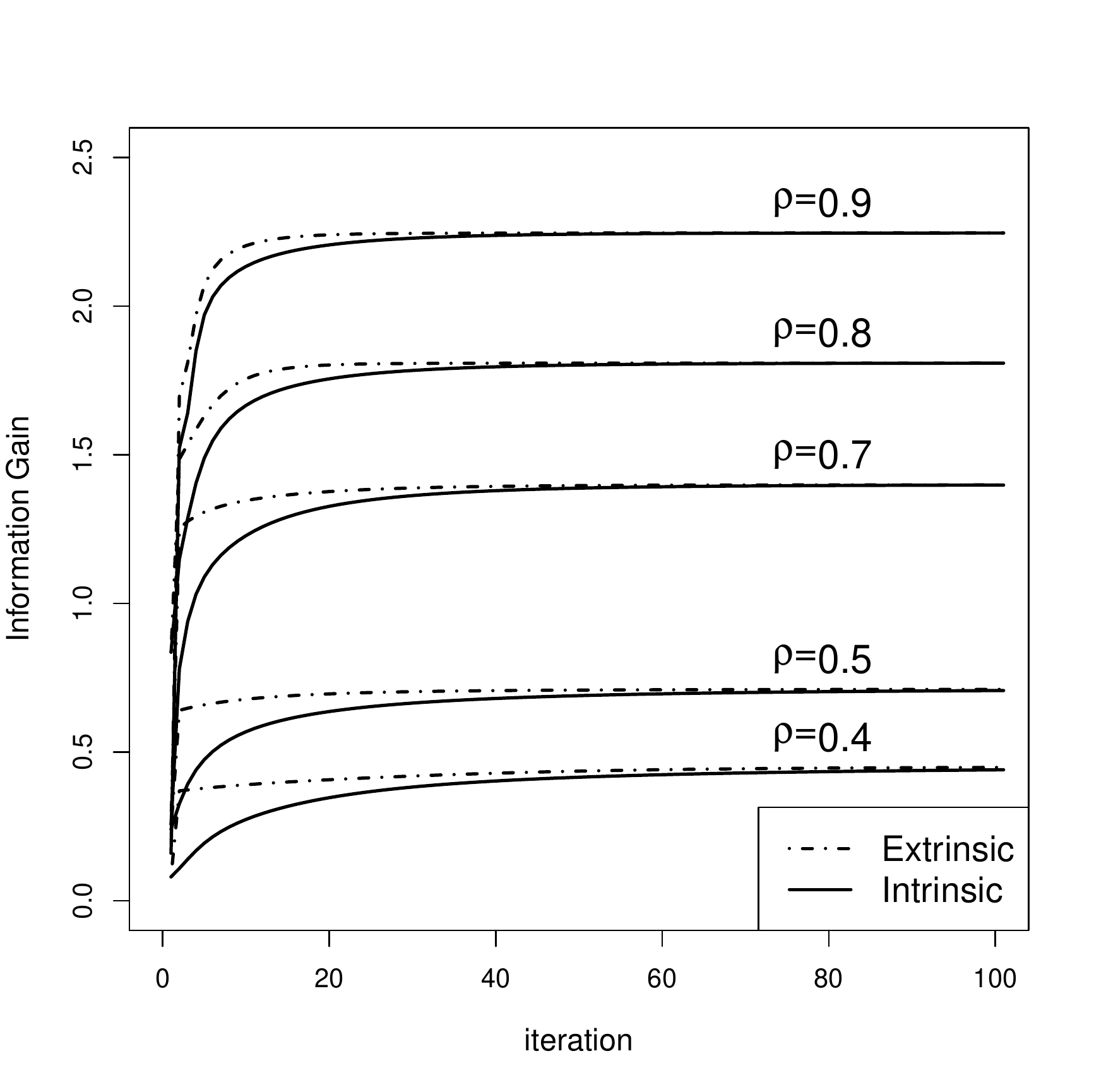}
\caption{A numerical study. The $x$-axis is the index for iteration and the $y$-axis is the information gain obtained at each iteration. The solid curve is for the intrinsic algorithm and the dashed curve is for the extrinsic algorithm.} % $\rho=0.5$, the dashed curve is for $\rho=0.7$ and the dot-dashed curve is for $\rho=0.9$.}
\label{fig:simu}
\end{figure}

 The results are shown in Fig.~\ref{fig:simu}. Here we set the step size $\delta_k=0.1$. The $x$-axis is the index for iterations and the $y$-axis gives the information gain for the secondary channel returned at step~$k$. First, it can be seen that, as $\rho$ increases, the information gain is increasing as well because the correlation between $\bm\theta$ and $\bm\phi$ is increasing. Next, it can also be seen that the performance of the two algorithms are quite comparable and both algorithms converge for each value of $\rho$. From our empirical evidence,  when the step size is constant, both algorithms would perform similarly, and in fact, the extrinsic algorithm converges slightly faster. For more complex problems, we could choose the optimal step size over a finite interval as suggested by the intrinsic algorithm in Section~\ref{sec:IntAlgo}. While, for extrinsic algorithm, such strategy for the optimal step size is not available.

\section{Discussion on Low-dimensional Channel Design}\label{Sec:DimDisc}
In the two-channel design problem considered in this paper, the number of measurements of the secondary channel, i.e., the number of rows of the channel matrix $\bm G$ is an important factor. Ideally we want $t$ to be as small as possible while keeping the information gain as large as possible. More measurements will generally bring more information. However, under the total power constraint, the information a channel carries is bounded and the upper bound may be attained by a small number of measurements. In fact, for a secondary channel with a $q$-dimensional input $\bm\phi$, a $q$-dimensional output $\bm y$ is sufficient to achieve the maximum information gain, which is a consequence of the following lemma. Here we assume that the measurement noise $\bm v$ in the secondary channel is white noise.
\begin{lemma}\label{lemma:dimreduc}
Suppose that the noise covariance $\bm Q_{\bm v\bm v}$ is proportional to the identity matrix. Then, for any channel matrix $\bm G\in\mathbb{R}^{t\times q}$ with rank $r$, there exists an $r$-dimensional secondary channel with the same noise variance that achieves the same information gain.
\end{lemma}

The proof is given in Appendix~\ref{Appen:lemma3pf}.

Since the maximum rank of $\bm G$ is $q$, Lemma~\ref{lemma:dimreduc} suggests that a $q$-dimensional $\bm y$ is sufficient to achieve the maximum information gain. Thus, we restrict our attention to the channel matrix with dimension $t\times q$ with $t\leq q$. In some cases, the power constraint will further reduce the dimension of $\bm y$ to $t<q$.  For instance, as shown in Example~1, the $5\times 5$ optimal channel matrices can have rank $5$, $4$, or $3$, and the dimension of $\bm y$ may be reduced correspondingly. 
Denote $\bm G_k^\ast$ the optimal channel matrix of dimension $k \times q$ for $k=1,\ldots, q$. %Let $r^\ast=\text{rank}(\bm G_q^\ast)$.
The optimal dimension of $\bm y$, denoted by $t^\ast$,  is defined as the smallest $k$ such that $D(\bm G_q^\ast)=D({\bm G}_k^\ast)$; that is, $t^\ast=\min\{k: D(\bm G_q^\ast)-D({\bm G}_k^\ast)=0 \}$.   Note that $D(\bm G^\ast_k)=D(\bm G_q^\ast)$ for any $k\geq t^\ast$, and $D(\bm G^\ast_k)<D(\bm G_q^\ast)$ for any $k< t^\ast$. In general, the values of $t^\ast$ is unknown since no analytical solution for $\bm G_k^\ast$ is available. From a  practical viewpoint, it is natural to approximate $t^\ast$ using the approximate optimal channel matrices. Here we consider the following approach to obtain an approximant of $t^\ast$.

For $k=1,\ldots,q$, obtain an approximate optimal channel matrix of dimension ${k\times q}$, denoted by  $\widehat{\bm G}_k^\ast$, using either the extrinsic or intrinsic algorithms.  Denote $\widehat{t}^\ast=\min\{k: D(\widehat{\bm G}_q^\ast)-D(\widehat{\bm G}_k^\ast)\leq c \}$, where $c$ is a predetermined threshold value, and $\widehat{t}^\ast$ is the proposed dimension of $\bm y$. The following example demonstrates this suggested strategy with more details.

\begin{example}
Consider a two-channel system \eqref{eqn:linsys} with $p=q=20$ and $s=10$. The channel matrix $\bm F\in\mathbb{R}^{10\times20}$ is randomly generated with Frobenius norm $1$. The noise covariances $\bm Q_{\bm u\bm u}=\bm Q_{\bm v\bm v}=\bm I_{20}$. The covariances $\bm Q_{\bm\theta\bm\theta}$ and $\bm Q_{\bm\phi\bm\phi}$ are randomly generated positive definite matrices. We consider two different correlation structures between $\bm\theta$ and $\bm\phi$: 1) $\bm Q_{\bm\phi\bm\phi|\bm\theta}=\bm I_{20}$ (analytical solution available); 2) $\bm Q_{\bm\phi\bm\phi|\bm\theta}$ is a banded matrix with $2$ on the main diagonal line and $0.2$ on the superdiagonal and subdiagonal lines (analytical solution not available).  The results are shown in Fig.~\ref{fig:dimG} and Fig.~\ref{fig:dimG2}, where the $x$-axis is $k$ ($k=1,\ldots, q$), number of rows of the secondary channel matrix $\bm G$, the $y$-axis on the left is the information gain for an $k$-dimensional secondary channel, and the $y$-axis on the right is the rank of the channel matrix with dimension $k\times q$.

 In the first scenario (Fig.~\ref{fig:dimG}), we obtain $\bm G_k^\ast$ for $k=1,\ldots, q$ analytically (shown in the left panel). It can be seen that the information gain remains constant for all $k\geq 4$. Therefore the optimal dimension is $t^\ast=4$. Moreover, one can see that the rank of all the optimal channel matrices $\bm G_k^\ast$ with $k\geq 4$ equal $4$, which may suggest that the optimal dimension $t^\ast$ equals the maximum rank of the optimal channel matrices. Therefore the curve for the rank of the optimal matrices can be used as an important guidance.
The extrinsic (the middle panel) and intrinsic (the right panel) algorithms are implemented, with the initial channel matrices randomly generated. Here we set the constant step size $\delta_k=0.1$. For both algorithms we get $\widehat{t}^\ast=t^\ast=4$ for $c=10^{-3}$, and so is the maximum rank.

 In the second scenario (Fig.~\ref{fig:dimG2}), we implement the extrinsic and intrinsic algorithms  to approximate the optimal channel matrix. Note that the solutions for $k=4$ and $k=5$ have similar information gain but different ranks. If the threshold value $c=10^{-3}$, we have $\widehat{t}^\ast=4$ in both algorithms, while the maximum rank equals $5$. Such difference may be caused by approximation error of the numerical algorithms.

\begin{figure}[htbp]
\centering
\includegraphics[totalheight=1.5in]{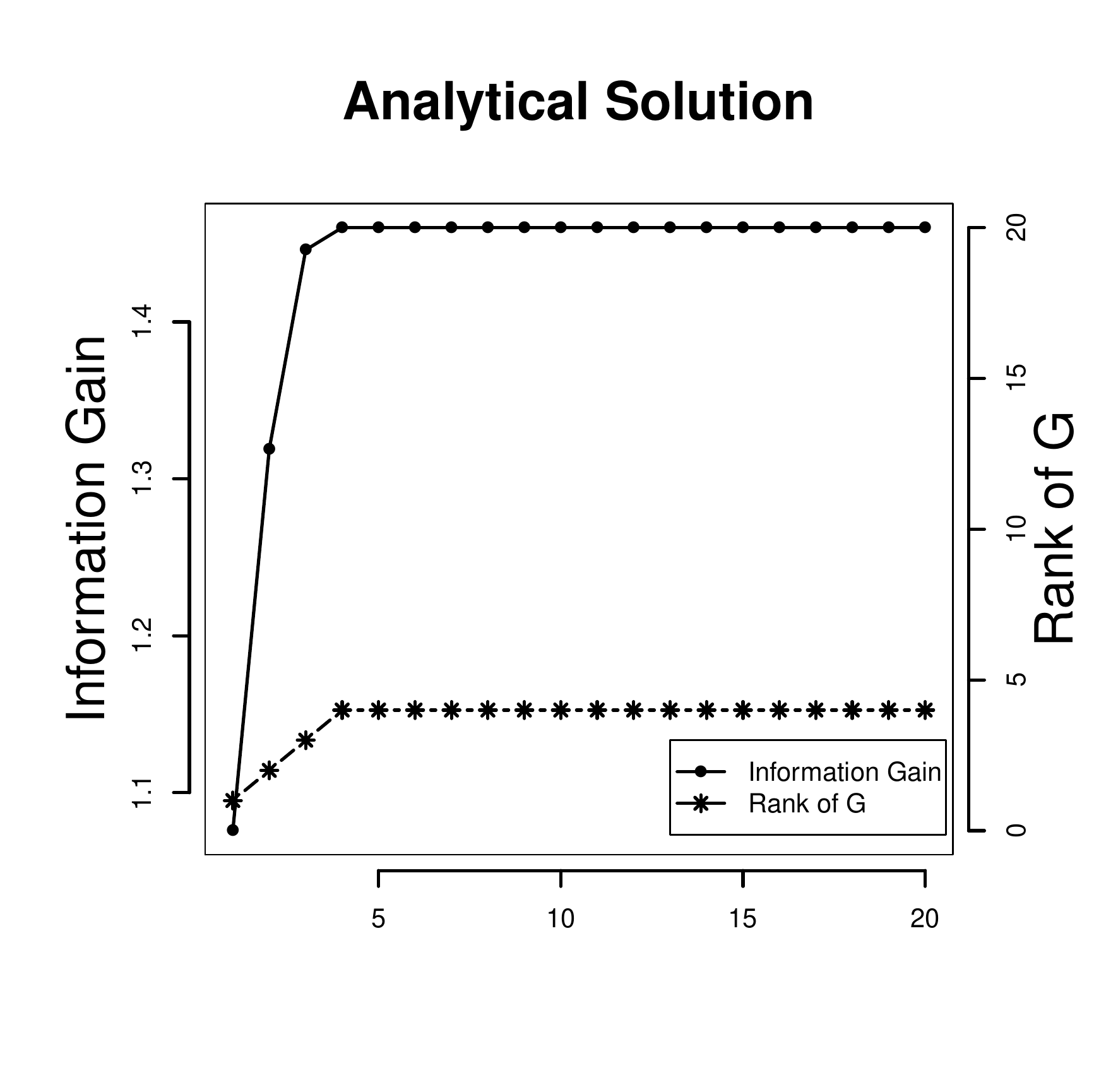}
\includegraphics[totalheight=1.5in]{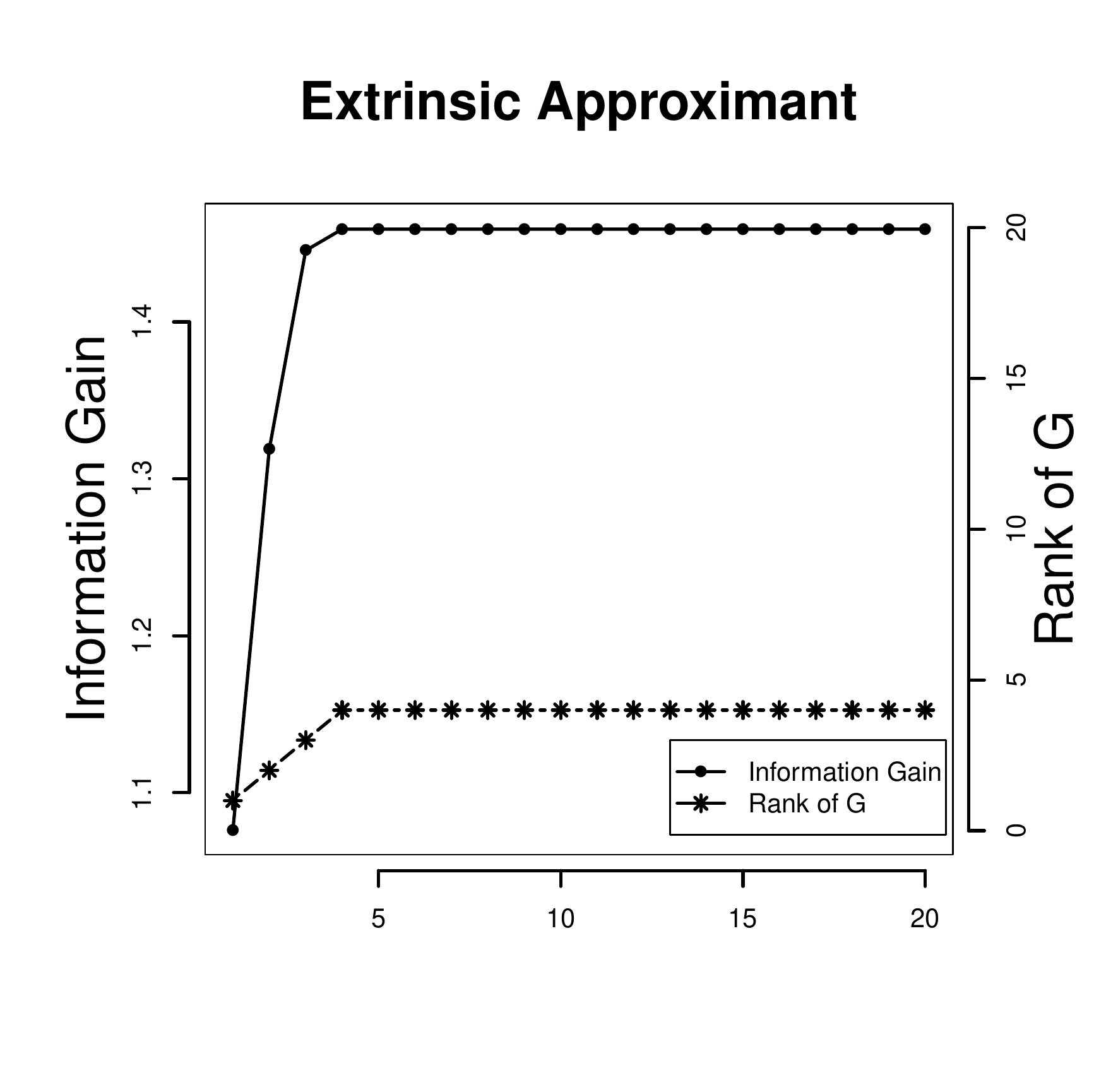}
\includegraphics[totalheight=1.5in]{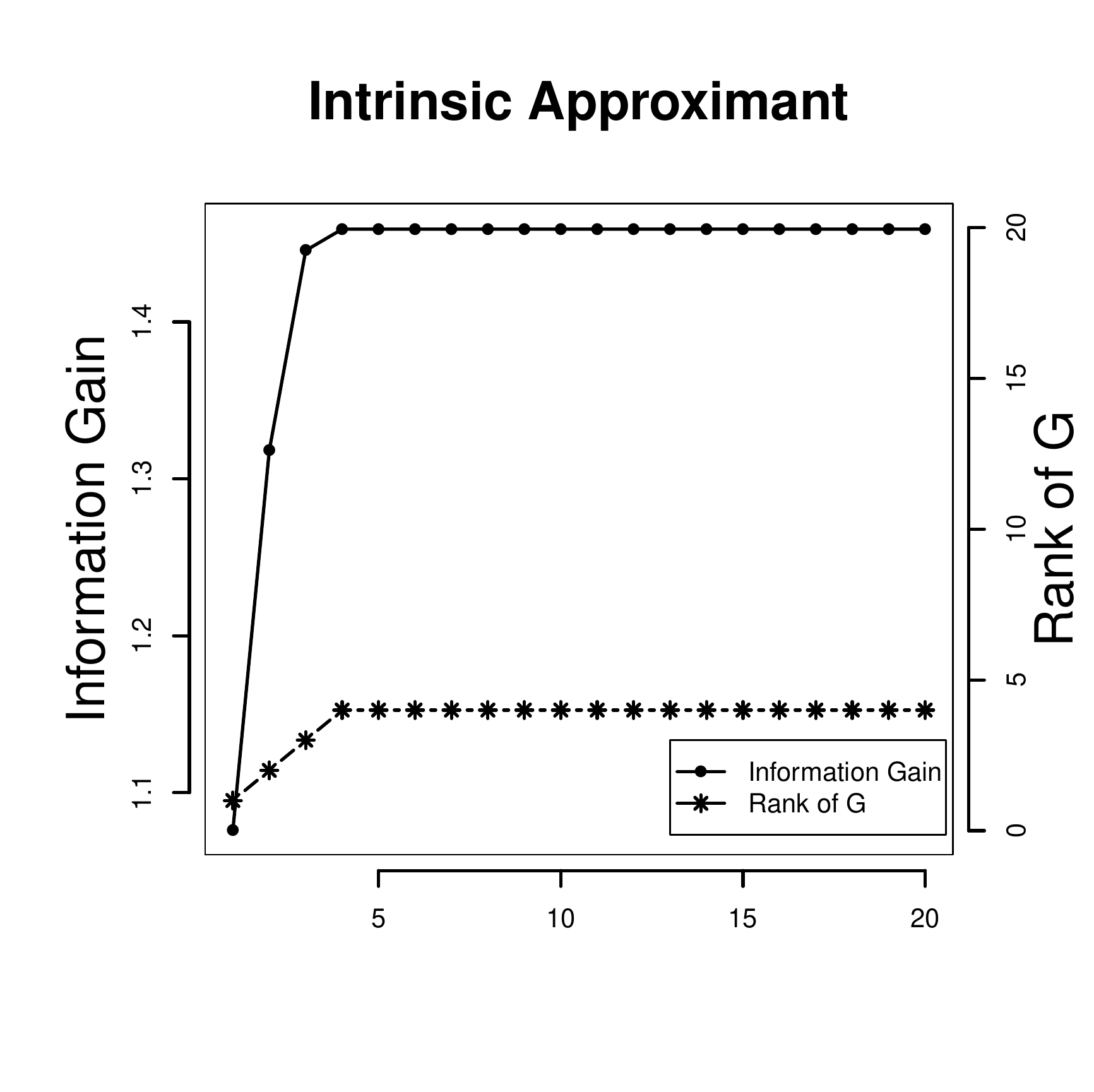}
\caption{Choice of number of rows of the secondary channel matrix $\bm G$. The three panels are associated with the channel matrices returned by the analytical solution (left), the extrinsic algorithm (middle) and the intrinsic algorithm (right), respectively.  In each panel, the $x$-axis indicates the number of rows of $\bm G$, the $y$-axis on the left is the information gain (solid line), and the $y$-axis on the right is the rank of the channel matrices (star dotted line). }\label{fig:dimG}
\end{figure}

\begin{figure}[htbp]
\centering
\includegraphics[totalheight=2in]{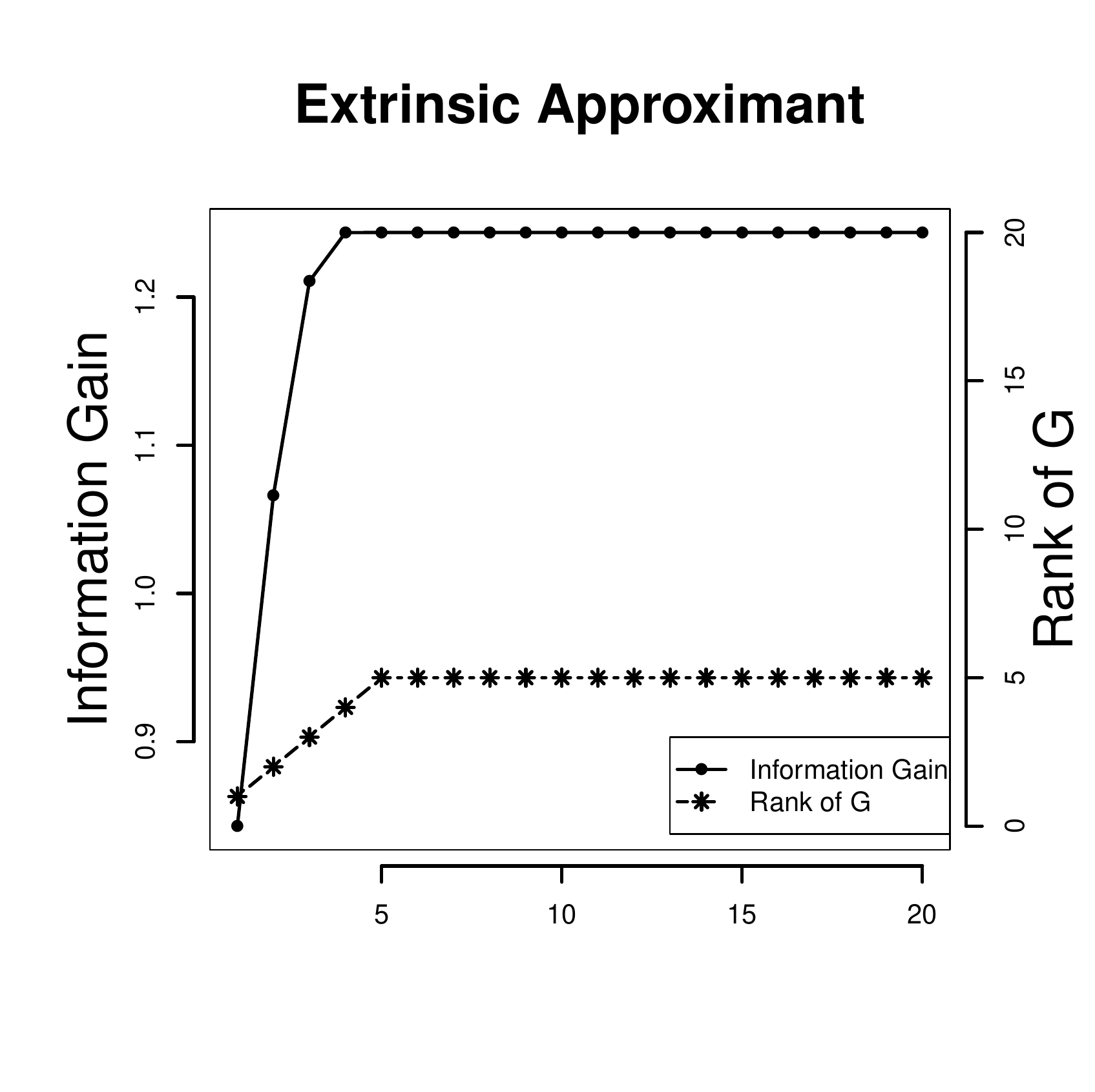}
\includegraphics[totalheight=2in]{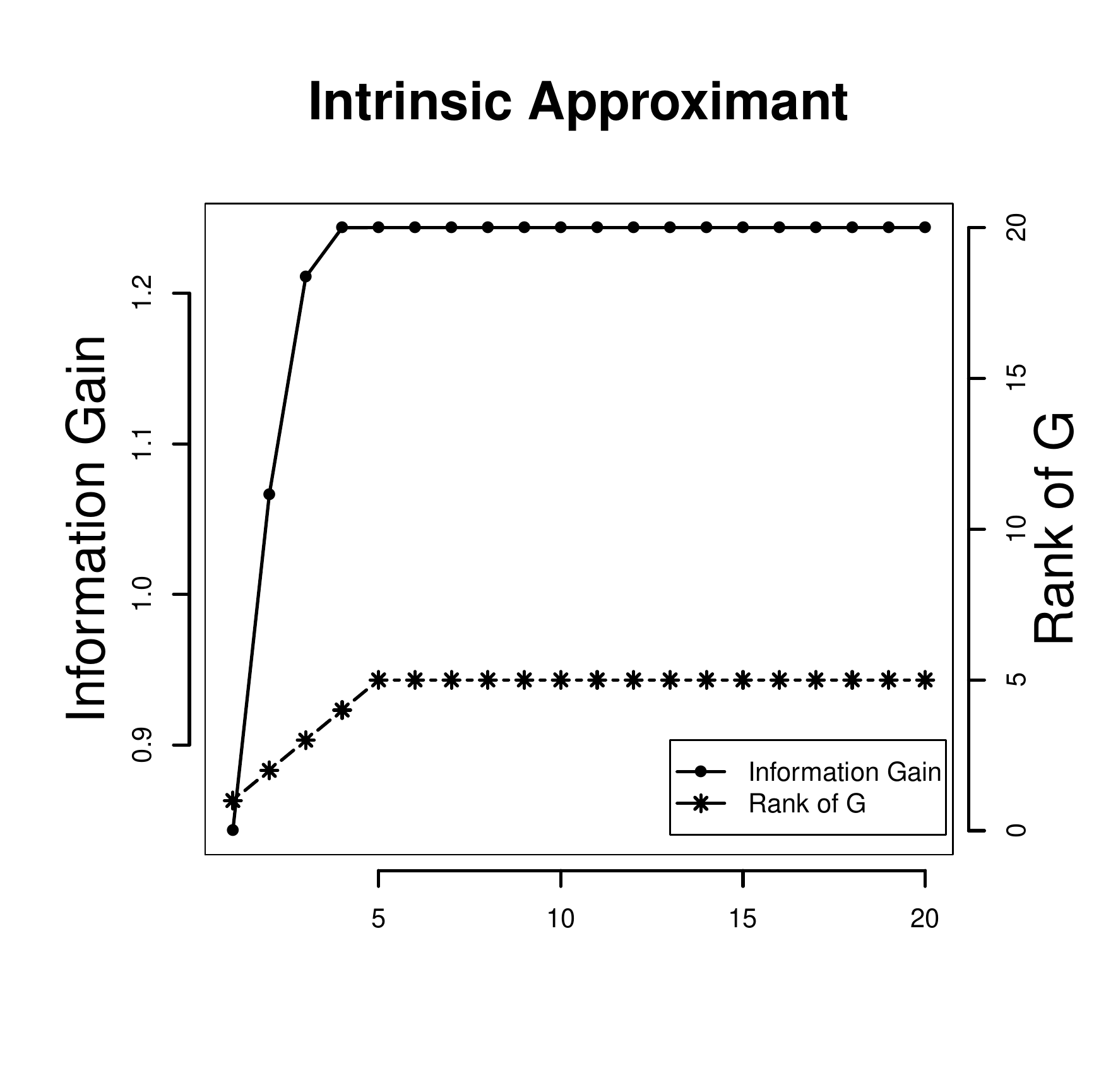}
\caption{Choice of number of rows of the secondary channel matrix $\bm G$. The two panels are associated with the channel matrices returned by the extrinsic algorithm (left) and the intrinsic algorithm (right), respectively.  In each panel, the $x$-axis indicates the number of rows of $\bm G$, the $y$-axis on the left is the information gain (solid line), and the $y$-axis on the right is the rank of the channel matrices (star dotted line). }\label{fig:dimG2}
\end{figure}
\end{example}

\section{Conclusions}\label{Sec:Conclu}

In this paper, we have studied the problem of fusing multiple sources of information. We have modeled the problem as a two-channel system where the signal in the primary channel is of interest, and the signal in the secondary channel is jointly distributed with the signal of interest. The objective is to design the secondary channel to maximize the information gain brought by fusing measurements from the primary and secondary channels. Based on the Gaussian distribution and linear channel assumptions, we obtain a closed-form expression of the information gain. When the input signals have a special covariance structure, we obtain an explicit solution for the optimal channel matrix, where the singular vectors are allocated to create non-interfering subchannels and the singular values solve a generalized water-filling problem. For general cases, we propose two gradient search algorithms, an extrinsic algorithm and an intrinsic algorithm to approximate the optimal channel matrix. Both algorithms can be extended to optimize other design criteria under a power constraint. With the designed secondary channel matrix,  combining the measurements of both channels achieves the best information gain.

\section*{Acknowledgements}
This work was supported in part by AFOSR contract FA 9550-10-1-0241, NSF grant CFF-1018472, and  DMS-1106975.

\appendix

\section{Proof of Lemma~\ref{lemma:phi}}\label{Appen:LemmaPhi}

By the matrix inversion lemma $\bm\Phi^T(\bm\Phi\bm\Phi^T+\bm\Sigma_{\bm v})^{-1}\bm\Phi=\bm I-(\bm I+\bm\Phi\bm\Sigma_{\bm v}^{-1}\bm\Phi^T)^{-1}$,   $D(\bm\Phi)$ may be rewritten as
\begin{align*}
D(\bm\Phi)&=\frac{1}{2}\log\det[\bm I+\bm\Sigma_{\bm\xi}-(\bm I+\bm\Phi\bm\Sigma_{\bm v}^{-1}\bm\Phi^T)^{-1}\bm\Sigma_{\bm\xi}]\\
&=\frac{1}{2}\log\det[\bm I+\bm\Sigma_{\bm\xi}]+\frac{1}{2}\log\det[\bm I-(\bm I+\bm\Phi\bm\Sigma_{\bm v}^{-1}\bm\Phi^T)^{-1}\bm\Sigma_{\bm\xi}(\bm I+\bm\Sigma_{\bm\xi})^{-1}]
\end{align*}
Define $\bm\Lambda:=\bm\Sigma_{\bm v}^{-1}$ a $t\times t$ diagonal matrix with diagonal elements $\lambda_{i}=\sigma_{\bm v,i}^{-2}$, and $\bm\Gamma:=\bm\Sigma_{\bm\xi}(\bm I+\bm\Sigma_{\bm\xi})^{-1}$ a $q\times q$ diagonal matrix with diagonal elements $\gamma_{i}=\sigma_{\bm\xi,i}^2/(1+\sigma_{\bm\xi,i}^2)$. Let $\rho$ be the rank of $\bm\Sigma_{\bm\xi}$. Then  $1>\gamma_1\geq\ldots\geq\gamma_{\rho}>0$ and $\gamma_{\rho+1}=\ldots=\gamma_{q}=0$.
The Lagrangian is
\begin{align}\label{Eqn:Langrdet}
L(\bm\Phi;\mu)&=\frac{1}{2}\log\det(\bm I_n-(\bm I_n+\bm\Phi^T\bm\Lambda\bm\Phi)^{-1}\bm\Gamma)+\mu (\operatorname{tr}(\bm\Phi\bm\Phi^T)- P)\nonumber\\
&+\frac{1}{2}\log\det[\bm I+\bm\Sigma_{\bm\xi}]
\end{align}
where $\mu$ is the Lagrangian multiplier.
The partial derivative of $L(\bm\Phi;\mu)$ with respect to the elements of $\bm\Phi$ is
\begin{align*}
\nabla_{\bm\Phi}L(\bm\Phi;\mu)=\bm\Lambda\bm\Phi(\bm I_n-\bm\Gamma+\bm\Phi^T\bm\Lambda\bm\Phi)^{-1}-
\bm\Lambda\bm\Phi(\bm I_n+\bm\Phi^T\bm\Lambda\bm\Phi)^{-1}+2\mu\bm\Phi.
\end{align*}
 Left multiply the gradient by $\bm\Lambda^{-1}$ and right multiply by $\bm\Phi^T$:
 \begin{align*}
-\bm\Phi(\bm I_n-\bm\Gamma+\bm\Phi^T\bm\Lambda\bm\Phi)^{-1}\bm\Phi^T+
\bm\Phi(\bm I_n+\bm\Phi^T\bm\Lambda\bm\Phi)^{-1}\bm\Phi^T=2\mu\bm\Lambda^{-1}\bm\Phi\bm\Phi^T
\end{align*}
Since the LHS is symmetric,  $\bm\Phi\bm\Phi^T\bm\Lambda^{-1}=\bm\Lambda^{-1}\bm\Phi\bm\Phi^T$. Therefore, when $\bm\Lambda$ has distinct diagonal elements, the symmetric matrix $\bm\Phi\bm\Phi^T$ must be diagonal.
Next we show that $\bm\Phi(\bm I_n-\bm\Gamma)^{-1}\bm\Phi^T$ is diagonal.  Notice that
\begin{align*}
&\bm\Phi(\bm I_n-\bm\Gamma+\bm\Phi^T\bm\Lambda\bm\Phi)^{-1}\bm\Phi^T=\bm\Lambda^{-1}-\bm\Lambda^{-1}(\bm\Lambda^{-1}+\bm\Phi(\bm I_n-\bm\Gamma)^{-1}\bm\Phi^T)\bm\Lambda^{-1}\\
&\bm\Phi(\bm I_n+\bm\Phi^T\bm\Lambda\bm\Phi)^{-1}\bm\Phi^T=\bm\Lambda^{-1}-\bm\Lambda^{-1}(\bm\Lambda^{-1}+\bm\Phi\bm\Phi^T)\bm\Lambda^{-1}
\end{align*}
Then, right multiply the gradient by  $\bm\Phi^T$:
\begin{align*}
\bm\Lambda^{-1}\bm\Phi(\bm I_n-\bm\Gamma)^{-1}\bm\Phi^T=\bm\Lambda^{-1}\bm\Phi\bm\Phi^T-\mu\bm\Phi\bm\Phi^T
\end{align*}
The RHS is symmetric since $\bm\Phi\bm\Phi^T$ is diagonal. Therefore we have $\bm\Lambda^{-1}\bm\Phi(\bm I_n-\bm\Gamma)^{-1}\bm\Phi^T=\bm\Phi(\bm I_n-\bm\Gamma)^{-1}\bm\Phi^T\bm\Lambda^{-1}$, which implies that $\bm\Phi(\bm I_n-\bm\Gamma)^{-1}\bm\Phi^T$ is diagonal.

Denote $\bm\Phi:=[\phi_{ij}]$. Given the fact that $\bm\Phi(\bm I_n-\bm\Gamma)^{-1}\bm\Phi^T$ and $\bm\Phi\bm\Phi^T$ are diagonal,  \eqref{Eqn:Langrdet} can be rewritten as
\begin{align*}
L(\bm\Phi;\mu)&=\frac{1}{2}\log\det(\bm I_m+\bm\Lambda\bm\Phi(\bm I_n-\bm\Gamma)^{-1}\bm\Phi^T)-\frac{1}{2}\log\det(\bm I_m+\bm\Lambda\bm\Phi\bm\Phi^T)\nonumber\\
&+\frac{1}{2}\log\det(\bm I_n-\bm\Gamma)+\frac{1}{2}\log\det[\bm I+\bm\Sigma_{\bm\xi}]+\mu (\operatorname{tr}(\bm\Phi\bm\Phi^T)- P)\nonumber\\
&=\sum_{i=1}^m\frac{1}{2}\log(1+\lambda_i\sum_{j=1}^n\frac{\phi_{ij}^2}{1-\gamma_j})-\sum_{i=1}^m\frac{1}{2}\log(1+\lambda_i\sum_{j=1}^n\phi_{ij}^2)\nonumber\\
&+\frac{1}{2}\log\det(\bm I_n-\bm\Gamma)+\frac{1}{2}\log\det[\bm I+\bm\Sigma_{\bm\xi}]+\mu (\sum_{i=1}^m\sum_{j=1}^n\phi_{ij}^2- P)
\end{align*}
Notice that $L(\bm\Phi;\mu)$ is quadratic in each $\phi_{ij}$. Therefore, we can assume WLOG $\phi_{ij}\geq0$.
The partial derivative of $L(\bm\Phi;\mu)$ w.r.t $\phi_{ij}$ is
\begin{align*}
\frac{\partial L(\bm\Phi;\mu) }{\partial \phi_{ij}}=\phi_{ij}\left[\frac{\lambda_{i}(1-\gamma_{j})^{-1}}{1+\sum_{j=1}^n \phi_{ij}^2(1-\gamma_{j})^{-1}}-
\frac{\lambda_{i}}{1+\sum_{j=1}^n \phi_{ij}^2}+2\mu\right]
\end{align*}
For $j>\rho$, we have $\gamma_{j}=0$, and  $L(\bm\Phi;\mu)$ is monotone decreasing in $\phi_{ij}$ since $\mu\leq0$. Hence for any minimizer $\bm\Phi$,  $\phi_{ij}=0$ for any $j>\rho$.

For the $i$th row, suppose that there exist two non-zero elements $\phi_{ij_1}$ and $\phi_{ij_2}$. Then the partial derivative $\frac{\partial L(\bm\Phi;\mu) }{\partial \phi_{ij_1}}=\frac{\partial L(\bm\Phi;\mu) }{\partial \phi_{ij_2}}=0$
yields
\begin{align*}
\frac{\lambda_{i}(1-\gamma_{j_1})^{-1}}{1+\sum_{j=1}^n \phi_{ij}^2(1-\gamma_{j})^{-1}}=\frac{\lambda_{i}(1-\gamma_{j_2})^{-1}}{1+\sum_{j=1}^n \phi_{ij}^2(1-\gamma_{j})^{-1}}
\end{align*}
which contradicts the assumption $\gamma_{j_1}\neq\gamma_{j_2}$. For the $j$th column, if there are two non-zero elements $\phi_{i_1j}$ and $\phi_{i_2j}$, then $\phi_{i_1k}=\phi_{i_2k}=0$ for any $k\neq j$ since each row of $\bm\Phi$ has at most one non-zero entry. Hence, $[\bm\Phi\bm\Phi^T]_{i_1i_2}=\sum_{k=1}^n \phi_{i_1k}\phi_{i_2k}=\phi_{i_1j}\phi_{i_2j}\neq0$, which contradicts diagonal $\bm\Phi\bm\Phi^T$.

\section{Proof of Theorem~\ref{thm:Gstar}}\label{Appen:Gstar}
Restricting the matrix $\bm\Phi$ within the class of matrices satisfying Lemma~\ref{lemma:phi}, $\bm\Phi$ can be written as $\bm\Phi=\bm\Pi_2\bm\Lambda\bm \Pi_1^T$  where $\bm\Pi_1\in\mathbb{R}^{q\times q}$ and $\bm\Pi_2\in\mathbb{R}^{t\times t}$ are permutation matrices and $\bm\Lambda$ is a $t \times q$ diagonal matrix with diagonal elements $\lambda_{11},\ldots,\lambda_{tt}$. The maximum information gain is taken over the permutations $\bm\Pi_1$ $\bm\Pi_2$ and $\lambda_{11}, \ldots, \lambda_{tt}$ subject to $\sum_{i=1}^t \lambda_{ii}^2\leq P$.

First of all, we show the optimal permutation matrices are $\bm\Pi_1=\bm I_q$  and $\bm\Pi_2=\bm I_t$. Denote
$$f(\bm\Pi_1, \bm\Pi_2)=\max D(\bm\Phi) \text{ subject to } \bm\Phi=\bm\Pi_2\bm\Lambda\bm \Pi_1^T \text{ and }\sum_{i=1}^t \lambda_{ii}^2\leq P.$$
The objective is to show
$$
f(\bm I_q, \bm I_t)\geq f(\bm\Pi_1, \bm\Pi_2)
$$
for all the possible permutations $\bm\Pi_1$ and $\bm\Pi_2$.

Let $\pi_1(i)$ be the index of the entry equal to unity in the $i$th column of $\bm\Pi_1$, and $\pi_2(i)$ the index of the unity entry in the $i$th column of $\bm\Pi_2$.  Then the information gain $D(\bm\Phi)$ can be written as
\begin{align*}
 D(\bm\Phi|\bm\Pi_1, \bm\Pi_2)%&=\frac{1}{2}\log\det(\bm I+\bm\Sigma^T(\bm\Sigma\bm\Sigma^T+\bm\Pi_2^T\bm\Sigma_{\bm v}\bm\Pi_2)^{-1}\bm\Sigma\bm\Pi_1^T\bm\Sigma_{\bm\xi}\bm\Pi_1)
 =\frac{1}{2}\sum_{i=1}^t\log\left(1+\frac{\sigma^2_{\bm\xi, \pi_1(i)}\lambda_{ii}^2}{\lambda_{ii}^2+\sigma^2_{\bm v, \pi_2(i)}}\right)
\end{align*}
It is easy to see that for any $i=1,\ldots, t$, one must have $$\sigma^2_{\bm\xi, \pi_1(i)} \geq \max\{\sigma^2_{\bm\xi, \pi_1(t+1)}, \ldots, \sigma^2_{\bm\xi, \pi_1(q)}\}.$$ Moreover, since the orders of $\{\pi_1(j)\}_{j=t +1}^q$ do not affect the value of $D(\bm\Phi)$, we can set WLOG $\pi_1(j)=j$ for $j=t +1,\ldots,q$. For $i=1,\ldots,t$, it can be seen that $\pi_1(i)$ and $\pi_2(i)$ are pairwise. Therefore, we can set WLOG that $\pi_{2}(i)=i$ for $i=1,\ldots,t$ and then search for the optimal permutation $\pi_1(i)$ to maximize
\begin{align}
 D(\bm\Phi|\bm\Pi_1, \bm I_t)=\frac{1}{2}\sum_{i=1}^t\log\left(1+\frac{\sigma^2_{\bm\xi, \pi_1(i)}\lambda_{ii}^2}{\lambda_{ii}^2+\sigma^2_{\bm v, i}}\right)
\end{align}
The proof that $\bm\Pi_1=\bm I_q$ is the optimal permutation matrix is similar to the proof for Theorem~2 in~\cite{Wang:Wang:Scharf:2013}. First prove the case $t=2$ and generalize the results to $t\geq 2$. The details are omitted.

Next, the objective is to  solve a  simpler optimization problem:
\begin{align}\label{Eqn:phistar}
\{\lambda_{ii}^\ast\}_{i=1}^t=\arg\max\frac{1}{2}\sum_{i=1}^t\log\left(1+\frac{\sigma^2_{\bm\xi, i}\lambda_{ii}^2}{\lambda_{ii}^2+\sigma^2_{\bm v, i}}\right) \text{ subject to} \quad \sum_{i=1}^t \lambda_{ii}^2\leq P.
\end{align}
The Lagrangian is
\begin{align}\label{Eqn:L0}
L(\lambda_{11}, \ldots, \lambda_{tt},;\mu)=\frac{1}{2}\sum_{i=1}^t\log\left(1+\frac{\sigma^2_{\bm\xi, \pi_1(i)}\lambda_{ii}^2}{\lambda_{ii}^2+\sigma^2_{\bm v, \pi_2(i)}}\right)-\mu (\sum_{i=1}^t\lambda_{ii}^2- P)
\end{align}
where $\mu\geq0$ is the Lagrange multiplier.
Setting the first derivative of $L$ w.r.t. $\lambda_{ii}$ equal to zero, we have either $\lambda_{ii}= 0$ or
\begin{align}\label{Eqn:L0deri}
\lambda_{ii}=\sqrt{\frac{b_i\left(-(2+a_i)+\sqrt{(2+a_i)^2-4(1+a_i)(1-a_i/(2\mu b_i))}\right)}{2(1+a_i)}}
\end{align}
where $a_i=\sigma^2_{\bm\xi, i}$ and $b_i=\sigma^2_{\bm v, i}$.
Equation \eqref{Eqn:L0deri} provides a feasible solution for ${\lambda}_{ii}$  when $\mu\leq a_i/(2b_i)$.  To see whether the solution is the maximizer for~\eqref{Eqn:phistar} , we check the Hessian matrix. The second derivative of $\bm L$ w.r.t. $\lambda_{ii}$ is
\begin{align}\label{Eqn:Hessian}
\frac{\partial ^2L(\bm\Phi;\mu)}{\partial\lambda_{ii}^2}=&-2\mu+\frac{a_ib_i}{(\lambda_{ii}^2+b_i)(\lambda_{ii}^2+b_i+a_i\lambda_{ii}^2)}\nonumber\\
&-\frac{2a_ib_i\lambda_{ii}^2((\lambda_{ii}^2+b_i)(2+a_i)+a_i\lambda_{ii}^2)}{(\lambda_{ii}^2+b_i)^2(\lambda_{ii}^2+b_i+a_i\lambda_{ii}^2)^2}
\end{align}
For $i=1\ldots,\kappa$, upon substituting \eqref{Eqn:L0deri},
$$
\frac{\partial ^2L(\bm\Phi;\mu)}{\partial\lambda_{ii}^2}=-\frac{8\mu^2\lambda_{ii}^2}{a_i}\sqrt{(2+a_i)^2-4(1+a_i)(1-\frac{a_i}{2\mu b_i})}.
$$
which is negative when $\mu<a_i/(2b_i)$.
For $\lambda_{ii}=0$,
$$
\left.\frac{\partial ^2L(\bm\Phi;\mu)}{\partial\lambda_{ii}^2}\right|_{\lambda_{ii}=0}=-2\mu+\frac{a_i}{b_i},
$$
 is negative when $\mu>a_i/(2b_i)$. Let $\kappa$ be the maximum integer such that $\mu<a_i/(2b_i)$ for $i=1,\ldots, \kappa$ with $\mu$ uniquely solves that $\sum_{i=1}^\kappa\lambda_{ii}^2=P$. Then, the maximizer of \eqref{Eqn:phistar} is $\lambda_{11}^\ast,\ldots,\lambda_{tt}^\ast$ where
\begin{align}\label{Eqn:phi}
\lambda_{ii}^{\ast 2}=\left\{ \begin{array}{ll} \frac{b_i\left(\sqrt{(2+a_i)^2-4(1+a_i)(1-a_i/(2\mu b_i))}-(2+a_i)\right)}{2(1+a_i)}, &\text{ for } i=1,\ldots,\kappa\\
0&\text{ for } i=\kappa+1,\ldots,t.
\end{array}
\right.
\end{align}

\section{Gradient of the Information Gain}\label{Appen:Gradient}
Applying the matrix inversion lemma yields
$$
\bm G^T(\bm G\bm Q_{\bm\phi\bm\phi|\bm\theta}\bm G^T+\bm Q_{\bm v\bm v})^{-1}\bm G=\bm Q_{\bm\phi\bm\phi|\bm\theta}^{-1}-\bm Q_{\bm\phi\bm\phi|\bm\theta}^{-1}(\bm G^T\bm Q_{\bm v\bm v}^{-1}\bm G+\bm Q_{\bm\phi\bm\phi|\bm\theta}^{-1})^{-1}\bm Q_{\bm\phi\bm\phi|\bm\theta}^{-1}.
$$
Therefore, the information gain $D(\bm G)$ satisfies
\begin{align*}
&2D(\bm G)\\
=&\log\det[\bm I_q+(\bm Q_{\bm\phi\bm\phi|\bm\theta}^{-1}-\bm Q_{\bm\phi\bm\phi|\bm\theta}^{-1}(\bm G^T\bm Q_{\bm v\bm v}^{-1}\bm G+\bm Q_{\bm\phi\bm\phi|\bm\theta}^{-1})^{-1}\bm Q_{\bm\phi\bm\phi|\bm\theta}^{-1})\bm M\bm Q_{\bm\theta\bm\theta|\bm x}\bm M^T]\\
=&\log\det[\bm I_q+ \bm Q_{\bm\phi\bm\phi|\bm\theta}^{-1}\bm M\bm Q_{\bm\theta\bm\theta|\bm x}\bm M^T]+\log\det[\bm I-\bm B(\bm G^T\bm Q_{\bm v\bm v}^{-1}\bm G+\bm Q_{\bm\phi\bm\phi|\bm\theta}^{-1})^{-1}]
\end{align*}
where $\bm B=\bm Q_{\bm\phi\bm\phi|\bm\theta}^{-1}\bm M\bm Q_{\bm\theta\bm\theta|\bm x}\bm M^T(\bm I_q+ \bm Q_{\bm\phi\bm\phi|\bm\theta}^{-1}\bm M\bm Q_{\bm\theta\bm\theta|\bm x}\bm M^T)^{-1}\bm Q_{\bm\phi\bm\phi|\bm\theta}^{-1}$.

Let $\bm J_{i,j}$ be a $t \times q$ matrix with value $1$ at element $(i,j)$ and $0$ elsewhere. From~\cite{Petersen:Pedersen:2006}, for a matrix $\bm X$, we have the partial derivatives
\begin{eqnarray*}
\partial \bm X^{-1}=-\bm X^{-1}(\partial \bm X)\bm X^{-1}, \quad \partial \log\det \bm X=\operatorname{tr}(\bm X^{-1}\partial \bm X).
\end{eqnarray*}
Let $\bm C=(\bm I_q-(\bm Q_{\bm\phi\bm\phi|\bm\theta}^{-1}+\bm G^T\bm Q_{\bm v\bm v}^{-1}\bm G)^{-1}\bm B)$. Then $D(\bm G)=\frac{1}{2}\log\det \bm C$ and we have
\begin{align*}
&\frac{\partial D}{\partial \bm G_{i,j}}=\frac{1}{2}\operatorname{tr}\{\bm C^{-1}\frac{\partial \bm C}{\partial \bm G_{ij}}\}=-\frac{1}{2}\operatorname{tr}\{\bm C^{-1}\frac{\partial(\bm Q_{\bm\phi\bm\phi|\bm\theta}^{-1}+\bm G^T\bm Q_{\bm v\bm v}^{-1}\bm G)^{-1}}{\partial \bm G_{ij}}\bm B\}\\
&=\frac{1}{2}\operatorname{tr}\{\bm C^{-1}(\bm Q_{\bm\phi\bm\phi|\bm\theta}^{-1}+\bm G^T\bm Q_{\bm v\bm v}^{-1}\bm G)^{-1}\frac{\partial (\bm Q_{\bm\phi\bm\phi|\bm\theta}^{-1}+\bm G^T\bm Q_{\bm v\bm v}^{-1}\bm G)}{\partial \bm G_{i,j}}\times \\
&(\bm Q_{\bm\phi\bm\phi|\bm\theta}^{-1}+\bm G^T\bm Q_{\bm v\bm v}^{-1}\bm G)^{-1}\bm B\}
=\frac{1}{2}\operatorname{tr}\{\bm C^{-1}(\bm Q_{\bm\phi\bm\phi|\bm\theta}^{-1}+\bm G^T\bm Q_{\bm v\bm v}^{-1}\bm G)^{-1}\times\\
&(\bm J_{i,j}^T\bm Q_{\bm v\bm v}^{-1}\bm G+\bm G^T\bm Q_{\bm v\bm v}^{-1}\bm J_{i,j})(\bm Q_{\bm\phi\bm\phi|\bm\theta}^{-1}+\bm G^T\bm Q_{\bm v\bm v}^{-1}\bm G)^{-1}\bm B\}\\
& =\left\{(\bm Q_{\bm\phi\bm\phi|\bm\theta}^{-1}+\bm G^T\bm Q_{\bm v\bm v}^{-1}\bm G)^{-1}\bm B\bm C^{-1}(\bm Q_{\bm\phi\bm\phi|\bm\theta}^{-1}+\bm G^T\bm Q_{\bm v\bm v}^{-1}\bm G)^{-1}\bm G^T\bm Q_{\bm v\bm v}^{-1}\right\}_{j,i}
 \end{align*}
 where the last equality follows from $\operatorname{tr}(\bm A\bm J_{ij})=\bm A_{j,i}=\operatorname{tr}(\bm J_{ij}^T\bm A^T)$. Hence, the gradient of function $D$ with respect to $\bm G$ is
\begin{eqnarray*}
\nabla_{\bm G} D=\bm Q_{\bm v\bm v}^{-1}\bm G[(\bm Q_{\bm\phi\bm\phi|\bm\theta}^{-1}+\bm G^T\bm Q_{\bm v\bm v}^{-1}\bm G)-\bm B]^{-1}\bm B(\bm Q_{\bm\phi\bm\phi|\bm\theta}^{-1}+\bm G^T\bm Q_{\bm v\bm v}^{-1}\bm G)^{-1}.
\end{eqnarray*}
\section{Proof of Lemma~2}\label{Appen:lemma3pf}
Suppose that the noise $\bm v$ has covariance $\bm Q_{\bm v\bm v}=\sigma_{\bm v}^2\bm I_t$. Then for any $t\times t$ orthogonal matrix $\bm U$,
\begin{align*}
&\quad D(\bm U\bm G)\\
&=\frac{1}{2}\log\det[\bm I_p+\bm M^T\bm G^T\bm U^T(\bm U\bm G\bm Q_{\bm\phi\bm\phi|\bm\theta}\bm G^T\bm U^T+\sigma_{\bm v}^2\bm I_t)^{-1}\bm U\bm G\bm M\bm Q_{\bm\theta\bm\theta|\bm x}]\\
&=\frac{1}{2}\log\det[\bm I_p+\bm M^T\bm G^T(\bm G\bm Q_{\bm\phi\bm\phi|\bm\theta}\bm G^T+\sigma_{\bm v}^2\bm I_t)^{-1}\bm G\bm M\bm Q_{\bm\theta\bm\theta|\bm x}]\\
&=D(\bm G)
\end{align*}
Therefore, the information gain $D(\bm G)$ is invariant to left unitary multiplication of $\bm G$. For any $\bm G\in\mathbb{R}^{t\times q}$ with rank $r$, $\bm G$ has the singular value decomposition $\bm G=\bm U\bm\Delta\bm V^T$ where  $\bm U$ and $\bm V$ are orthogonal matrices, and  $\bm \Delta\in\mathbb{R}^{t\times q}$ is a diagonal matrix with diagonal elements $\bm\Delta_{1,1}\geq\ldots\geq\bm\Delta_{r,r}>0$ and $\bm\Delta_{i,i}=0$ for any $i\geq r$. By the invariance property, we can assume WOLG that $\bm U=\bm I_t$.
Let $\widetilde{\bm G}=\operatorname{Diag}(\bm\Delta_{1,1},\ldots,\bm\Delta_{r,r})\bm V_r^T\in\mathbb{R}^{r\times q}$ where $\bm V_r\in\mathbb{R}^{q \times r}$ contains the first $r$ columns of $\bm V$. It can be seen that $\bm G=[\widetilde{\bm G}^T, \bm 0_{q \times (t-r)}]^T$ and $\operatorname{tr}(\bm G\bm G^T)=\operatorname{tr}(\widetilde{\bm G}\widetilde{\bm G}^T)$. Moreover, one can easily check that
\begin{align}\label{Eqn:appen4}
D(\bm G)=\frac{1}{2}\log\det[\bm I_p+\bm M^T\widetilde{\bm G}^T(\widetilde{\bm G}\bm Q_{\bm\phi\bm\phi|\bm\theta}\widetilde{\bm G}^T+\sigma_{\bm v}^2\bm I_r)^{-1}\widetilde{\bm G}\bm M\bm Q_{\bm\theta\bm\theta|\bm x}].
\end{align}
The RHS of~\eqref{Eqn:appen4} is the information gain brought by a $r$-dimensional channel $\widetilde{\bm y}$ as
$$\widetilde{\bm y}=\widetilde{\bm G}\bm\phi+\widetilde{\bm v}$$
where $\widetilde{\bm v}$ is $r$-dimensional white noise with variance $\sigma_{\bm v}^2$.
The new channel $\widetilde{\bm y}$ brings the same information gain, that is
$I(\bm\theta; \bm x, \bm y)-I(\bm\theta; \bm x)=I(\bm\theta; \bm x,\widetilde{\bm y})-I(\bm\theta; \bm x)$.

\bibliographystyle{IEEEbib}

\end{document}